\documentclass[11pt]{article}

\usepackage{amsmath}
\usepackage{graphicx}
\usepackage{indentfirst}
\usepackage{amssymb}
\usepackage{cite}
\usepackage{color}
\usepackage{subfigure}
\usepackage{varwidth}
\usepackage{subfigure}
\usepackage[colorlinks=true, linkcolor=red, citecolor=blue, urlcolor=magenta]{hyperref}
\usepackage{xcolor}

\begin{document}
\title{Unveiling Inner Shadows and Polarization Signatures of Rotating Einstein–Gauss–Bonnet Black Holes}

\date{}
\maketitle

\begin{center}
\author{Bing-Bing Chen}$^{a}$\footnote{chenbingbing323512@163.com}
\author{Chen-Yu Yang}$^{b}$\footnote{chenyu\_yang2024@163.com}
\author{Deyou Chen}$^{c}$\footnote{Corresponding author: deyouchen@hotmail.com}
\author{Ke-Jian He}$^{b}$\footnote{kjhe94@163.com}

\vskip 0.25in
$^{a}$\it{School of Mathematics, Physics and Statistics, Sichuan Minzu College, Kangding 626001, People's Republic of China}\\
$^{b}$\it{Department of Mechanics, Chongqing Jiaotong University, Chongqing 400074, People's Republic of China}\\
$^{c}$\it{School of Science, Xihua University, Chengdu 610039, People’s
Republic of China}\\
\end{center}
\vskip 0.6in
{\abstract
{Based on the backward ray-tracing method, this paper numerically investigates the shadow and polarization images of rotating Einstein–Gauss–Bonnet (EGB) black hole within the framework of a thin disk model. We systematically analyze the effects of the main model parameters and the observation inclination angle $\theta_o$ on both types of images. The results show that, as an intrinsic property of the black hole, the inner shadow undergoes significant deformation with increasing $\theta_o$. The increase of the GB coupling constant $\xi$ only reduces the size of the inner shadow, while the spin parameter $a$ does not alter its size but also its shape. And, the photon ring is more sensitive to variations in $\theta_o$, while it is less affected by $\xi$ and $a$. 
For polarization images, the influence of $\xi$ on the polarization intensity is generally consistent with that observed in the accretion disk images. However, the polarization direction near the region of the inner shadow and photon ring changes significantly with $\xi$. This feature can provide an additional and effective observational tool for extracting information about the spacetime structure in Einstein--Gauss--Bonnet (EGB) gravity.
Finally, we conclude that, compared to previous reliance on either accretion disk or polarization images alone, the simultaneous combination and synergistic analysis of both can more profoundly reveal the optical properties of rotating EGB black holes, providing a stronger theoretical basis for identifying such black holes through future high-resolution observations.
}}

\thispagestyle{empty}
\newpage
\setcounter{page}{1}


\section{Introduction}
Since 2019, the Event Horizon Telescope (EHT) international collaboration has released images of the supermassive black holes at the center of the radio galaxy M87* and of our Galaxy, SgrA*~\cite{EventHorizonTelescope:2019dse,EventHorizonTelescope:2022wkp}, marking a new era in black hole physics. Both images exhibit similar structures: a bright, asymmetric ring surrounding a central region of diminished intensity. According to the theoretical framework of General Relativity (GR), the former corresponds to the photon ring, while the latter represents the black hole shadow. The shadow arises because light rays passing near the black hole are captured by its strong gravitational field and cannot reach the observer, thereby forming a dark region on the image plane~\cite{Cunha:2018acu}. The shadow image of a black hole encodes valuable information about the spacetime geometry, accretion processes, and the surrounding matter distribution~\cite{Falcke:1999pj,Zeng:2022pvb,Zeng:2021mok}. Consequently, it provides an unprecedented opportunity to test GR in the strong-field regime and to probe the complex plasma environment around compact objects~\cite{Gralla:2019xty}. Hence, constructing theoretical models of black hole shadows consistent with observational images has become one of the frontiers of contemporary black hole physics~\cite{Cardoso:2019rvt}.

Supermassive black holes accrete hot, magnetized plasma, forming luminous accretion disks. On the other hand, for highly spinning black holes, electromagnetic energy can drive relativistic jets~\cite{Blandford:1977ds}. In theory, both the accretion disk and the jet can act as light sources that generate black hole shadow images. In 1973, Shakura and Sunyaev first proposed the geometrically thin and optically thick standard accretion disk model~\cite{Shakura:1972te}. Subsequently, in 1974, by taking into account the relativistic effects of matter motion near black holes~\cite{Lightman:1974sm}, the Shapiro–Lightman–Eardley model was introduced, assuming that the accretion disk is both geometrically and optically thin. In 1979, Luminet demonstrated that the Schwarzschild black hole appears perfectly circular for any observation inclination angle~\cite{Luminet:1979nyg}. Later, the shadow images of black holes with Keplerian accretion disks in Kerr spacetime were also investigated~\cite{Beckwith:2004ae}. In recent years, spherical accretion models~\cite{Narayan:2019imo,Zeng:2020dco,Heydari-Fard:2023ent}, optically and geometrically thin disk models~\cite{Li:2024ctu,Zeng:2021dlj,Li:2021riw,Yang:2025byw,Yang:2024nin}, and geometrically thick disk models~\cite{Zhang:2024jrw,Zhang:2024lsf,Gjorgjieski:2024csb} have all been extensively discussed. The observational achievements of the EHT have brought black hole shadow research into a new stage. Based on the EHT data, numerous studies have focused on constraining the parameters of various black hole models~\cite{Kuang:2022ojj,Li:2025awg}. Meanwhile, related topics such as hot spot images~\cite{Huang:2024wpj}, polarized images~\cite{Lee:2022rtg,Hou:2024qqo,Chen:2024cxi}, jet images~\cite{Zhang:2024lsf}, and boson star images~\cite{He:2025qmq,Gjorgjieski:2024csb,Zeng:2025nmu,Li:2025awg} have also attracted widespread interest.

In numerical simulations of black hole shadows, polarization images play a crucial role. By comparing theoretically obtained polarization images with observational data, one can gain deeper insights into the astrophysical properties of accretion flows~\cite{Zhu:2022amy}. For a Schwarzschild black hole background, the EHT collaboration successfully reproduced the observed polarization image of M87* by employing the analytical approximation of light rays derived by Beloborodov~\cite{Beloborodov:2002mr,EventHorizonTelescope:2021btj}. This approach was later extended to the Kerr black hole. Gelles \textit{et al.} constructed a simplified model with an equatorial emission source, producing the corresponding polarization image~\cite{Gelles:2021kti}. These studies suggest that polarization characteristics are governed not only by magnetic fields but also by black hole spacetime parameters and the observation inclination angle. Beyond the Schwarzschild and Kerr black holes, the polarization images of other black hole types and horizonless ultra-compact objects have also been widely investigated, and obtained a series of important results~\cite{Qin:2021xvx,Shi:2024bpm,Deliyski:2023gik}. 

Motivated by these developments, it is therefore important to further explore the shadow and polarization images of black holes in alternative gravitational scenarios. Such studies may provide new theoretical templates for future high-resolution observations and offer additional possibilities for testing the nature of strong gravity. Among various modified gravity theories, Einstein–Gauss–Bonnet (EGB) gravity provides a particularly well-motivated framework to investigate gravitational phenomena beyond general relativity. 
It is widely recognized that the uniqueness of the Einstein field equations is established on the basis of the Lovelock theorem~\cite{Lovelock:1972vz}. However, in higher-dimensional spacetimes ($D > 4$), the Einstein--Hilbert action is no longer unique. An important example is the EGB gravity, whose theoretical origin lies in the heterotic string theory~\cite{Lanczos:1938sf,Lovelock:1971yv}. The EGB gravity has attracted extensive attention because it not only represents the low-energy limit of string theory~\cite{Zwiebach:1985uq} but also gives rise to ghost-free, nontrivial gravitational self-interactions~\cite{Nojiri:2018ouv}. Within this framework, one can explore fundamental problems of gravitation in a broader theoretical context than that of GR~\cite{Wiltshire:1985us}. The spherically symmetric and static black hole solution in EGB theory was first proposed by Boulware and Deser~\cite{Boulware:1985wk}, and subsequently, a variety of interesting black hole solutions have been obtained for various sources~\cite{Cai:2001dz,Neupane:2003vz,Dehghani:2002wn,torii2005spacetime}. For the spherically symmetric black hole, one has analyzed the observable features of EGB black holes under the spherical accretion model and explored the influence of the GB parameter on the black hole shadow\cite{Zeng:2020dco}. However, the observable features of rotating EGB black holes, particularly their shadow and polarization images, still await systematic investigation. To this end, we strategically adopt an optically thin and geometrically thin accretion disk as the emission source model, aiming to reveal the unique spacetime characteristics of EGB gravity through precise numerical simulations of the shadow images of black hole. Furthermore, the polarization structure of rotating EGB black holes has not yet been explored. An in-depth study of their polarization images, combined with accretion disk imaging, may not only provide theoretical grounds for testing the validity of EGB gravity theory but also offer key observational criteria for distinguishing EGB gravity from general relativity.
This paper aims to provide a systematic theoretical analysis for understanding the observational features of rotating EGB black holes through the above systematic investigation.

The structure of this paper is organized as follows. In Sec.~\ref{sec2}, we briefly review the exact solution of the rotating EGB black hole and present the equations of null geodesics. Section~\ref{sec3} introduces the ray-tracing techniques and the camera projection method. In Secs.~\ref{sec4} and~\ref{sec5}, we investigate the shadow and polarization images of the rotating EGB black hole illuminated by a thin accretion disk. Finally, Sec.~\ref{sec6} provides the conclusion and discussion. Throughout this work, we adopt the geometrized unit system with $c = G = 1$, where $c$ and $G$ denote the speed of light in vacuum and the gravitational constant, respectively.

\section{Review of Rotating EGB Black Holes}\label{sec2}
In higher-dimensional ($D > 4$) spacetimes, one has pointed out in pioneering works that the action of Einstein-Gauss-Bonnet (EGB) gravity can include quadratic correction terms constructed from curvature tensor invariants, with the action taking the form\cite{Lanczos:1938sf,Lovelock:1971yv,Kumar:2020owy},
\begin{equation}
S_{\mathrm{EGB}} = \frac{1}{16\pi G_D} 
\int d^D x \, \sqrt{-g} \, \left( \mathcal{L}_{\mathrm{EH}} + \alpha \mathcal{L}_{\mathrm{GB}} \right),
\end{equation}
where
\begin{equation}
\mathcal{L}_{\mathrm{EH}} = R, 
\qquad 
\mathcal{L}_{\mathrm{GB}} = R^{\mu\nu\rho\sigma} R_{\mu\nu\rho\sigma} 
- 4 R^{\mu\nu} R_{\mu\nu} + R^2.
\end{equation}
Here, $g$ is the determinant of the metric $g_{\mu\nu}$, and $\alpha$ is the Gauss--Bonnet (GB) coupling constant, 
which can be identified as the inverse string tension and is positive-definite.

In the Boyer–Lindquist (BL) coordinate system, the metric of a rotating EGB black hole is given by~\cite{Azreg-Ainou:2014pra,Azreg-Ainou:2014aqa}
\begin{align}
	ds^{2} &= -\left(\frac{\Delta - a^{2}\sin^{2}\theta}{\Sigma}\right) dt^{2}
	+ \frac{\Sigma}{\Delta} dr^{2}
	- 2a\sin^{2}\theta\left(1 - \frac{\Delta - a^{2}\sin^{2}\theta}{\Sigma}\right) dt\, d\varphi \label{eq:ma} \\
	&\quad + \Sigma\, d\theta^{2}
	+ \sin^{2}\theta \left[\Sigma + a^{2}\sin^{2}\theta\left(2 - \frac{\Delta - a^{2}\sin^{2}\theta}{\Sigma}\right)\right] d\varphi^{2}, \nonumber
\end{align}
where
\begin{equation}
	\Delta = r^{2} + a^{2} + \frac{r^{4}}{32\pi\xi}
	\left[1 - \left(1 + \frac{128\pi\xi M}{r^{3}}\right)^{\frac{1}{2}}\right],
	\qquad
	\Sigma = r^{2} + a^{2}\cos^{2}\theta.
\end{equation}
Here, $M$ and $a$ denote the black hole mass and spin parameter, respectively, while $\xi$ represents the Gauss--Bonnet (GB) coupling constant, which is considered to be positive definite and corresponds to the inverse string tension. In the limit $\xi \to 0$ or $r \to \infty$, the metric~(\ref{eq:ma}) reduces to that of the Kerr black hole~\cite{Kerr:1963ud}. Similar to the Kerr case, the spacetime admits two linearly independent Killing vector fields, $(\frac{\partial}{\partial t})^{\mu}$ and $(\frac{\partial}{\partial \varphi})^{\mu}$, which correspond to time-translation and rotational symmetries, respectively.

The horizons and coordinate singularities of the rotating EGB black hole described by the metric~(\ref{eq:ma}) can be determined by the roots of
\begin{equation}
	g^{\mu\nu} \partial_\mu r \, \partial_\nu r = g^{rr} = \Delta = 0.
	\label{eq:delta}
\end{equation}
For different values of $M$, $a$, and $\xi$, Eq.~(\ref{eq:delta}) admits three possibilities: two distinct real positive roots, two coincident real positive roots, or no real positive roots. These correspond to a non-extremal black hole, an extremal black hole, and a non-black-hole configuration, respectively. In this paper, we focus on the non-extremal case, where Eq.~(\ref{eq:delta}) possesses two positive real roots, denoted by the inner (Cauchy) horizon radius $r_-$ and the outer (event) horizon radius $r_+$, satisfying $r_- \leq r_+$. 
An observer whose angular momentum with respect to spatial infinity vanishes is defined as a zero–angular–momentum observer (ZAMO). The ZAMO co-rotates with the black hole due to the frame-dragging effect~\cite{Chandrasekhar:1985kt}. For the metric~(\ref{eq:ma}), the angular velocity $\omega$ of the ZAMO is
\begin{equation}
	\omega = \frac{d\varphi}{dt} = -\frac{g_{t\varphi}}{g_{\varphi\varphi}}
	= \frac{a\left(r^{2} + a^{2} - \Delta\right)}{(r^{2} + a^{2})^{2} - \Delta a^{2} \sin^{2}\theta}.
\end{equation}
The angular velocity $\omega$ increases monotonically as the radial coordinate $r$ decreases, reaching its maximum value at the event horizon $r = r_+$,
\begin{equation}
	\Omega = \omega\big|_{r=r_+} = \frac{a}{r_+^{2} + a^{2}}
	= \frac{32\pi\xi a}{r_+^{4}\left[-1 + \left(1 + \frac{128\pi\xi M}{r_+^{3}}\right)^{1/2}\right]},
\end{equation}
where $\Omega$ can be interpreted as the angular velocity of the black hole. 

The null geodesic equations accurately describe the trajectories of photons in the vicinity of a black hole. By applying the Hamilton–Jacobi equation, the geodesic equations for the metric~(\ref{eq:ma}) can be derived as
\begin{align}
	\Sigma\frac{dt}{d\tau} &= \frac{r^{2} + a^{2}}{\Delta}\left[E(r^{2} + a^{2}) - aL\right] - a(aE\sin^{2}\theta - L), \label{eq:ge1}\\
	\Sigma\frac{dr}{d\tau} &= \pm\sqrt{R(r)},\\
	\Sigma\frac{d\theta}{d\tau} &= \pm\sqrt{\Theta(\theta)},\\
	\Sigma\frac{d\varphi}{d\tau} &= \frac{a}{\Delta}\left[E(r^{2} + a^{2}) - aL\right] - \left(aE - \frac{L}{\sin^{2}\theta}\right), \label{eq:ge4}
\end{align}
where $\tau$ is the affine parameter along the geodesic, and the energy and angular momentum of photon are $E$ and $L$, respectively, as well as $K$ denotes the Carter constant~\cite{Carter:1968rr}. This are the equations of motion for photons, which can be used for the study of black hole shadows. However, this paper aims to perform numerical simulations of black hole accretion disk images. To this end, we adopt the Hamiltonian form of the photon geodesic equation for numerical solving to achieve imaging.

\section{The Imaging Method of a black hole}\label{sec3}
In this section, we briefly introduce the ray-tracing techniques~\cite{Cunha:2015yba}, which form the foundation of thin-disk imaging. The key to image construction lies in distinguishing between the light rays that reach the observer and those captured by the event horizon, the latter forming the dark region of the shadow on the projection plane. Within the domain of the ZAMO, a local orthonormal tetrad can be established as
\begin{equation}\label{e32}
{e_{0}}= \frac{g_{\phi\phi}\partial_{t}-g_{t\phi}\partial_{\phi}}{\sqrt{g_{\phi\phi}(g_{t\phi}^{2}-g_{\phi\phi}g_{tt})}}, \quad
e_{1}=-\frac{\partial_{r}}{\sqrt{g_{rr}}},\quad
e_{2}= \frac{\partial_{\theta}}{\sqrt{g_{\theta\theta}}},\quad
e_{3}= - \frac{\partial_{\phi}}{\sqrt{g_{\phi\phi}}}.
\end{equation}
where $e_0$ denotes the timelike vector corresponding to the observer’s four-velocity, and $e_1$ points toward the black hole center. To describe the trajectories of photons as viewed by the observer, celestial coordinates $(\alpha, \beta)$ are introduced. Suppose the observer is located at point $O$, and the vector $\overrightarrow{OA}$ represents the tangent vector of a null geodesic passing through $O$. The celestial sphere is a three-dimensional sphere centered at $O$ with radius $|\overrightarrow{OA}|$. The celestial coordinates $\alpha$ and $\beta$ represent the angles between $\overrightarrow{OA}$ and the basis vectors $e_1$ and $e_2$, respectively. For a geodesic $S(\tau) = (t(\tau), r(\tau), \theta(\tau), \varphi(\tau))$, its tangent vector can be expressed as a linear combination of $e_\mu$,
\begin{equation}
	\dot{S} = |\overrightarrow{OA}|(-e_0 + \cos\alpha\, e_1 + \sin\beta\cos\alpha\, e_2 + \sin\alpha\sin\beta\, e_3),
\end{equation}
where the overdot denotes differentiation with respect to the affine parameter $\tau$, and the negative sign indicates that the tangent vector is directed toward the past. In the ZAMO frame, the photon’s four-momentum is $p_{(\mu)} = p_{\nu} e_{(\mu)}^{\nu}$, where $p_{\nu}$ is the four-momentum in the BL coordinate system. The relationship between the photon’s four-momentum and the celestial coordinates $(\alpha, \beta)$ is given by~\cite{Hu:2020usx}
\begin{equation}
	\cos\alpha = \frac{p^{(1)}}{p^{(0)}}, \quad \tan\beta = \frac{p^{(3)}}{p^{(2)}}.
\end{equation}
On the projection screen, a Cartesian coordinate system $(x, y)$ is further introduced, which relates to the celestial coordinates through
\begin{equation}
	x = -2\tan\frac{\alpha}{2}\sin\beta, \quad y = -2\tan\frac{\alpha}{2}\cos\beta.
	\label{eq:proco}
\end{equation}

Next, we discuss the camera projection method. In this work, we adopt a wide-field fisheye camera model, where the field-of-view angle $\gamma_{\mathrm{fov}}$. For convenience in computation, the half-angle $\gamma_{\mathrm{fov}}/2$ is taken along both the Cartesian $x$ and $y$ directions, defining a square screen with side length
\begin{equation}
	L = 2\left|\overrightarrow{OA}\right|\tan\frac{\gamma_{\mathrm{fov}}}{2}.
\end{equation}
In the ray-tracing techniques, pixel mapping is required. Hence, the imaging plane is divided into $n \times n$ pixels, each of which has a side length
\begin{equation}
	l = \frac{L}{n} = \frac{2}{n}\left|\overrightarrow{OA}\right|\tan\frac{\gamma_{\mathrm{fov}}}{2}.
\end{equation}
The center of each pixel is labeled by coordinates $(i, j)$, where the lower-left pixel corresponds to $(1, 1)$ and the upper-right pixel to $(n, n)$, with $i, j \in [1, n]$. The relationship between the Cartesian coordinates $(x, y)$ and the pixel coordinates $(i, j)$ is given by
\begin{equation}
	x = l\left(i - \frac{n + 1}{2}\right), \quad 
	y = l\left(j - \frac{n + 1}{2}\right).
	\label{eq:proco2}
\end{equation}
By comparing Eqs.~(\ref{eq:proco}) and~(\ref{eq:proco2}), the relation between the pixel coordinates $(i, j)$ and the celestial coordinates $(\alpha, \beta)$ can be expressed as
\begin{align}
	\tan\frac{\alpha}{2} &= \frac{1}{n}\tan\left(\frac{\gamma_{\mathrm{fov}}}{2}\right)
	\left[\left(i - \frac{n + 1}{2}\right)^{2} + \left(j - \frac{n + 1}{2}\right)^{2}\right]^{1/2}, \nonumber \\
	\tan\beta &= \frac{2j - (n + 1)}{2i - (n + 1)}.
\end{align}
Based on the aforementioned ray-tracing and projection techniques, we continue to construct the images of a rotating Einstein–Gauss–Bonnet (EGB) black hole within the framework of an accretion disk model in next section.

\section{Thin Disk Images of EGB black hole}\label{sec4}
\subsection{Accretion Disk Model and The Total Intensity }
In realistic astrophysical scenarios, black holes are typically surrounded by accretion disks. Therefore, the study of the shadow images of black holes illuminated by accretion disks is of significant importance. In this work, we assume that the accretion disk lies in the equatorial plane and is both optically and geometrically thin, while the observer is located sufficiently far away from the black hole. Under these assumptions, the accretion disk can be regarded as being composed of a free electrically neutral plasma moving along timelike geodesics. The interior of a black hole acts as a one-way membrane region, and for a stationary black hole, the event horizon corresponds to the beginning of this membrane. Matter within the accretion disk that approaches the black hole too closely will inevitably fall into the event horizon, whereas matter located sufficiently far away can maintain stable circular orbits around the black hole. Following Ref.~\cite{Hou:2022eev}, we introduce the inner stable circular orbit (ISCO) as the boundary separating these two regions. Beyond the ISCO, the particles in the accretion disk move along Keplerian orbits, while inside the ISCO, they follow the critical plunging orbits and accelerate toward the event horizon. This dynamical behavior has been confirmed by astronomical observations~\cite{Chael:2021rjo}.

The location of the ISCO is determined by the effective potential of the massive particle. For a massive, electrically neutral particle located in the equatorial plane, the effective potential can be expressed as
\begin{equation}
	\mathcal{V}_e(r) = 1 + g^{tt}\mathcal{E}^{2} + g^{\varphi\varphi}\mathcal{L}^{2} - 2g^{t\varphi}\mathcal{E}\mathcal{L}.
\end{equation}
Here, $\mathcal{E}$ and $\mathcal{L}$ denote the specific energy and specific angular momentum of the electrically neutral particle, respectively. Similar to $E$ and $L$, the quantities $\mathcal{E}$ and $\mathcal{L}$ are conserved along the geodesics
\begin{align}
	\mathcal{E} &= -\frac{g_{tt} + g_{t\varphi}\omega}{\left(-g_{tt} - 2g_{t\varphi}\omega - g_{\varphi\varphi}\omega^{2}\right)^{1/2}},\\
	\mathcal{L} &= \frac{g_{t\varphi} + g_{\varphi\varphi}\omega}{\left(-g_{tt} - 2g_{t\varphi}\omega - g_{\varphi\varphi}\omega^{2}\right)^{1/2}},
\end{align}
where $\omega$ represents the angular velocity given by
\begin{equation}
	\omega = \frac{d\varphi}{dt} \equiv
	\frac{\dfrac{\partial g_{t\varphi}}{\partial r} +
		\left(\dfrac{\partial^{2} g_{t\varphi}}{\partial r^{2}} -
		\dfrac{\partial g_{tt}}{\partial r}
		\dfrac{\partial g_{\varphi\varphi}}{\partial r}\right)^{1/2}}
	{\dfrac{\partial g_{\varphi\varphi}}{\partial r}}.
	\label{eq:av}
\end{equation}
The ISCO radius $r_I$ satisfies the following conditions
\begin{equation}
	\left.\mathcal{V}_e(r)\right|_{r=r_I} = 0, \quad
	\left.\frac{\partial \mathcal{V}_e(r)}{\partial r}\right|_{r=r_I} = 0, \quad
	\left.\frac{\partial^{2} \mathcal{V}_e(r)}{\partial r^{2}}\right|_{r=r_I} = 0.
\end{equation}
Within the accretion disk, particles located in the region $r_+ < r \leq r_I$ plunge into the event horizon along the critical plunging orbits, and their four-velocity is given by
\begin{align}
	u_\mathrm{in}^{t} &= -g^{tt}\mathcal{E}_I + g^{t\varphi}\mathcal{L}_I,\\
	u_\mathrm{in}^{r} &= -\left(-\frac{1 + g_{tt}(u_\mathrm{in}^{t})^{2} + 2g_{t\varphi}u_\mathrm{in}^{t}u_\mathrm{in}^{\varphi} + g_{\varphi\varphi}(u_\mathrm{in}^{\varphi})^{2}}{g_{rr}}\right)^{1/2},\\
	u_\mathrm{in}^{\theta} &= 0,\\
	u_\mathrm{in}^{\varphi} &= -g^{t\varphi}\mathcal{E}_I + g^{\varphi\varphi}\mathcal{L}_I,
\end{align}
where $\mathcal{E}_I$ and $\mathcal{L}_I$ denote the conserved quantities of the particle at $r = r_I$. The negative sign in $u_\mathrm{in}^{r}$ indicates that the motion of the particle is directed toward the black hole. For the region $r > r_I$, the particles in the accretion disk move along Keplerian orbits, and their four-velocity can be written as
\begin{equation}
	u_\mathrm{out}^{\mu} =
	\left(\frac{1}{-g_{tt} - 2g_{t\varphi}\omega - g_{\varphi\varphi}\omega^{2}}\right)^{1/2}
	(1, 0, 0, \omega).
\end{equation}

On the other hand, due to the gravitational lensing effect, photons emitted from the accretion disk may cross the black hole’s equatorial plane multiple times before reaching the observer. Each intersection enhances the brightness of the shadow image~\cite{Zhang:2023bzv}. The image formed by the first intersection is referred to as the direct image, while that formed by the second intersection is called the lensed image. When the number of intersections exceeds two, the resulting images are collectively termed the higher-order images. It should be noted that this phenomenon persists even when photon reflection occurs or when the thickness of the accretion disk varies.

In general, the radial coordinates of the intersections between a light ray and the equatorial plane are not identical and are denoted as $r_n$ ($n = 1, 2, \dots, N$), where $N$ represents the maximum number of intersections. When a light ray interacts with the accretion disk, its intensity changes due to the emission and absorption of photons. Neglecting reflection effects, this process can be described by the radiative transfer equation
\begin{equation}
	\frac{d}{d\tau}\left(\frac{\tilde{I}_\nu}{\nu^3}\right)=\frac{\tilde{E}_\nu-\tilde{A}_\nu \tilde{I}_\nu}{\nu^2},\label{ligstr}
\end{equation}
where $\tau$ is the affine parameter along the null geodesic, and $\tilde{I}_{\nu}$, $\tilde{E}_{\nu}$, and $\tilde{A}_{\nu}$ represent the specific intensity, emissivity, and absorption coefficient at frequency $\nu$, respectively. When photon emission and absorption are absent, $\tilde{I}_{\nu}/\nu^3$ remains conserved along the geodesic. Under the thin-disk approximation, the accretion disk lies in the equatorial plane, such that $\tilde{E}_\nu=\tilde{A}_\nu=0$ outside the plane. In this case, the total intensity at each position on the observer’s screen is given by
\begin{equation}
\tilde{I}_{o}=\sum\limits_{n=1}^{N_{\max}}g_n^3 \tilde{E}_n.\label{eq:io}
\end{equation}
Considering that the EHT observes black hole images at a wavelength of $1.3~\mathrm{mm}$ ($230~\mathrm{GHz}$), the emissivity of the thin disk $\tilde{E}_{\nu}$ is modeled as a second-order polynomial in logarithmic space, expressed as
\begin{equation}
	\tilde{E}_{\nu}(r)=e^{\left(-\frac12k^2-2k\right)},\quad k  =\log\frac r{r_+}.
\end{equation}
Here, $g_n\equiv\nu_0/\nu_n$ is the redshift factor, where $\nu_0$ denotes the photon frequency measured on the observer’s screen and $\nu_n$ denotes the frequency measured in the locally stationary frame comoving with the accretion disk. Since the accretion disk is composed of an electrically neutral plasma moving along timelike geodesics with conserved quantities $\mathcal{E}$ and $\mathcal{L}$, for $r>r_{I}$ the redshift factor can be written as
\begin{equation}
	g_n^{\mathrm{out}}=\left.\frac{c_2\left(1-c_1\frac{p_\varphi}{p_t}\right)}{c_3\left(1+\omega\frac{p_\varphi}{p_t}\right)}\right|_{r=r_n},\quad r>r_{I},
\end{equation}
where
\begin{equation}
	c_1=\frac{g_{t\varphi}}{g_{\varphi\varphi}},\quad
	c_2=\left(\frac{-g_{\varphi\varphi}}{g_{tt}g_{\varphi\varphi}-g_{t\varphi}^{2}}\right)^{\frac{1}{2}},\quad
	c_3=\left(\frac{-1}{g_{tt}+2g_{t\varphi}\omega+g_{\varphi\varphi}\omega^{2}}\right)^{\frac{1}{2}},
\end{equation}
and $\omega$ is calculated from Eq.~(\ref{eq:av}). When $r<r_{I}$, the accreting matter plunges into the event horizon along the plunging orbits with a radial velocity $u_\mathrm{in}^{r}$, and the corresponding redshift factor is given by
\begin{equation}
	g_{n}^{\mathrm{in}}=-\left.\left[u_\mathrm{in}^{r}\frac{p_r}{p_t}-\mathcal{E}_{I}\left(g^{tt}-g^{t\varphi}\frac{p_\varphi}{p_t}\right)+\mathcal{L}_{I}\left(g^{\varphi\varphi}\frac{p_\varphi}{p_t}+g^{t\varphi}\right)\right]^{-1}\right|_{r=r_n},\quad r<r_{I}.
\end{equation}
Therefore, once the redshift factor and the accretion disk model are specified, the observed image of the EGB black hole in the thin disk model can be further computed based on Eq. (\ref{eq:io}).

\subsection{Numerical Results}
This subsection presents the numerical results for the shadow images of a rotating EGB black hole under the thin accretion disk model. For the spacetime described by Eq.~(\ref{eq:ma}), we set the black hole mass to $M=1$ and focus on examining the effects of the GB coupling constant $\xi$, the spin parameter $a$, and the observation inclination angle $\theta_o$ on the shadow images. In the numerical calculations, the observer is placed at a distance of $r_o = 500M = 500$, and the inclination angles are chosen as $\theta_o = 0^\circ$, $17^\circ$, and $70^\circ$. A prograde accretion disk configuration is adopted, in which the direction of the disk’s rotation is the same as that of the black hole spin.

\begin{figure}[!h]
	\centering 
	\subfigure[$\xi=0.005,\theta_o=0^\circ$]{\includegraphics[scale=0.4]{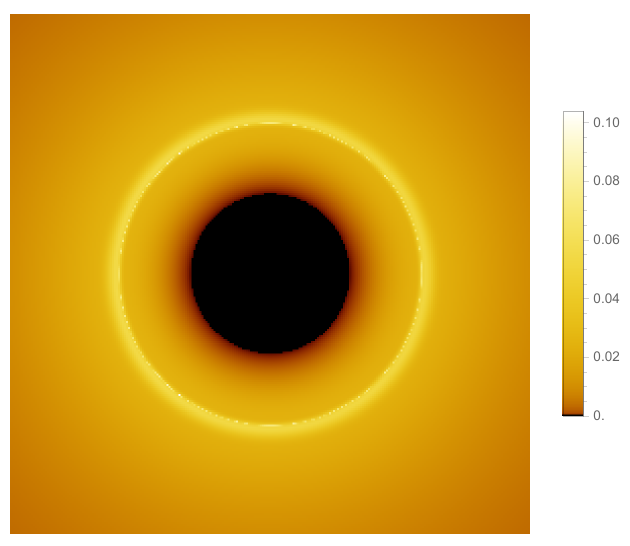}}
	\subfigure[$\xi=0.005,\theta_o=17^\circ$]{\includegraphics[scale=0.4]{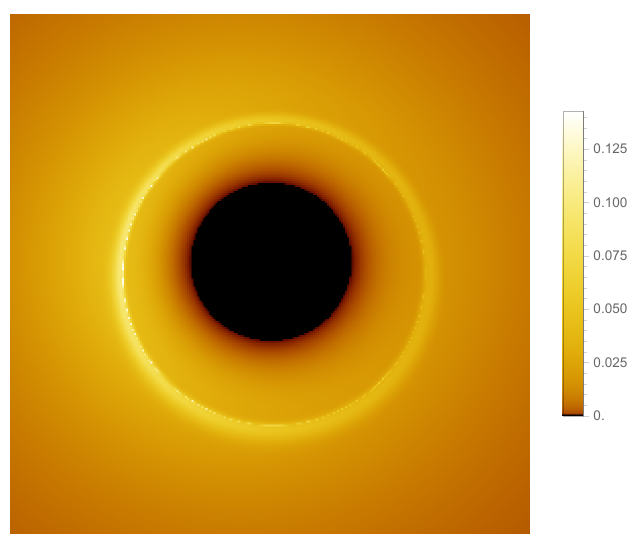}}
	\subfigure[$\xi=0.005,\theta_o=70^\circ$]{\includegraphics[scale=0.4]{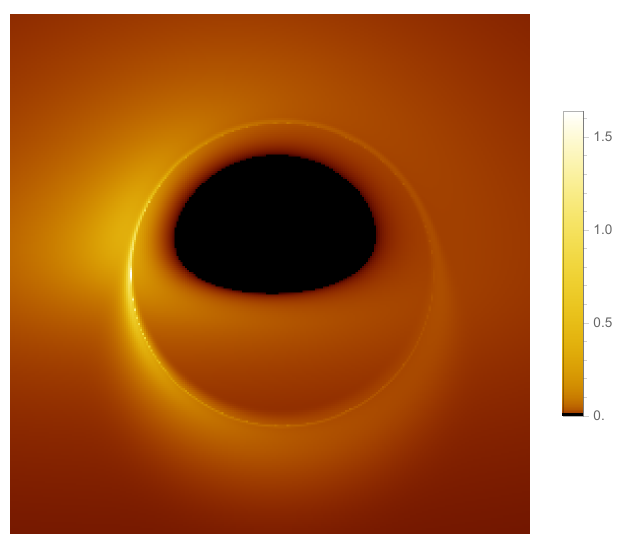}}
	
	\subfigure[$\xi=0.01,\theta_o=0^\circ$]{\includegraphics[scale=0.4]{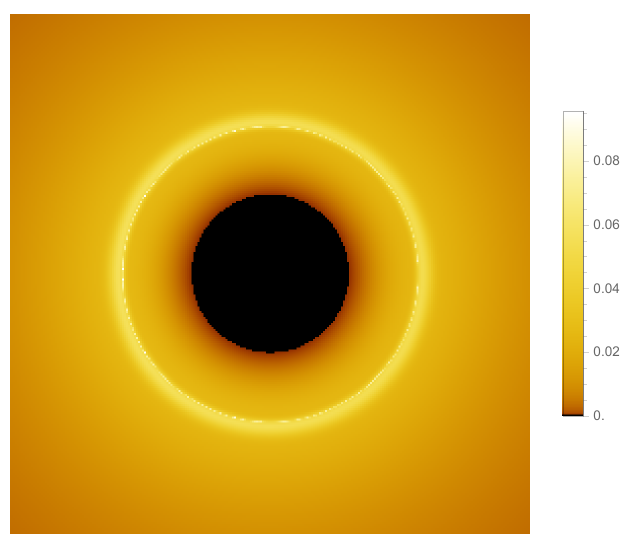}}
	\subfigure[$\xi=0.01,\theta_o=17^\circ$]{\includegraphics[scale=0.4]{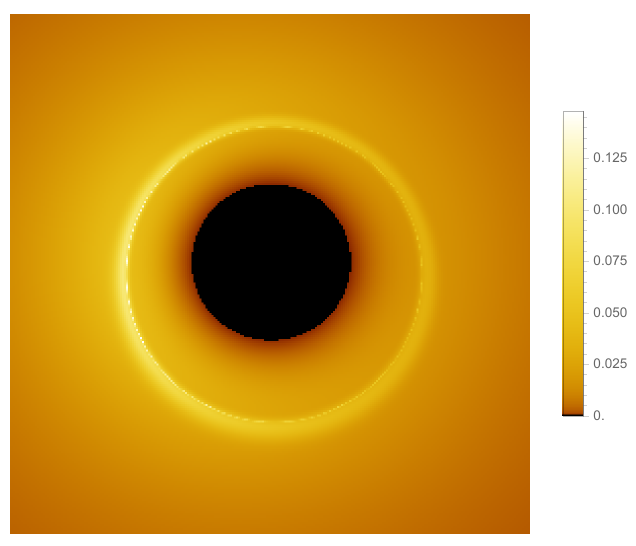}}
	\subfigure[$\xi=0.01,\theta_o=70^\circ$]{\includegraphics[scale=0.4]{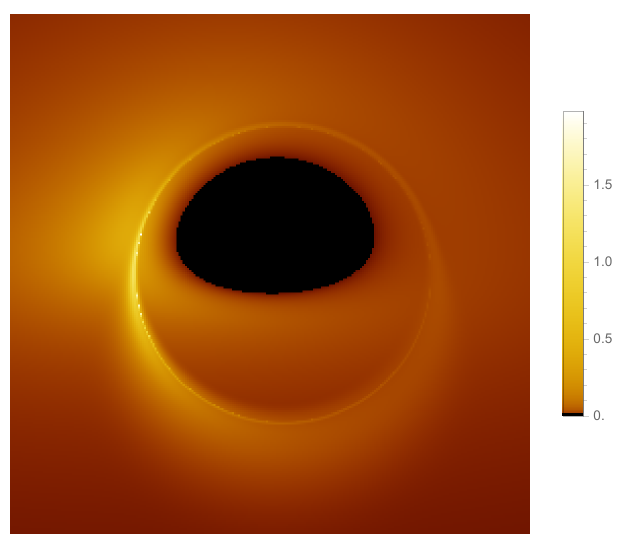}}
	
	\subfigure[$\xi=0.015,\theta_o=0^\circ$]{\includegraphics[scale=0.4]{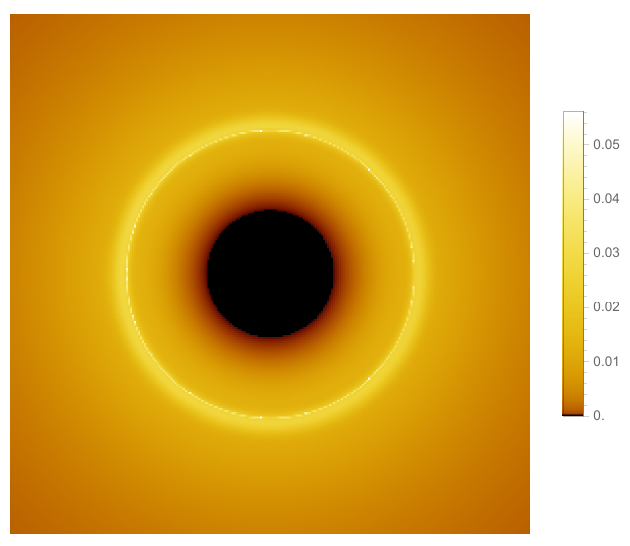}}
	\subfigure[$\xi=0.015,\theta_o=17^\circ$]{\includegraphics[scale=0.4]{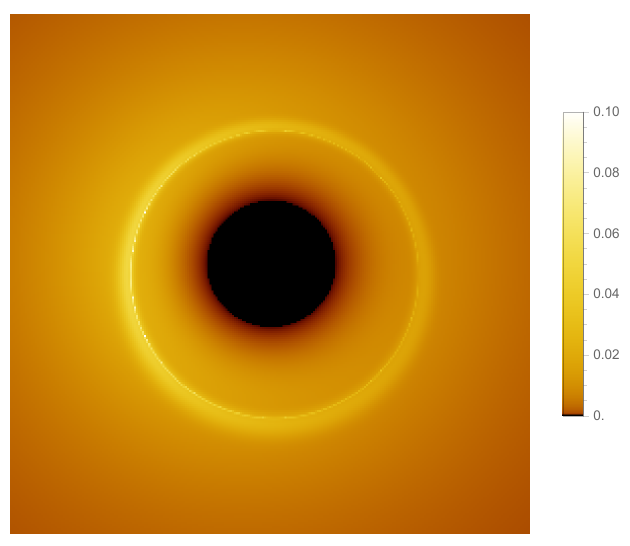}}
	\subfigure[$\xi=0.015,\theta_o=70^\circ$]{\includegraphics[scale=0.4]{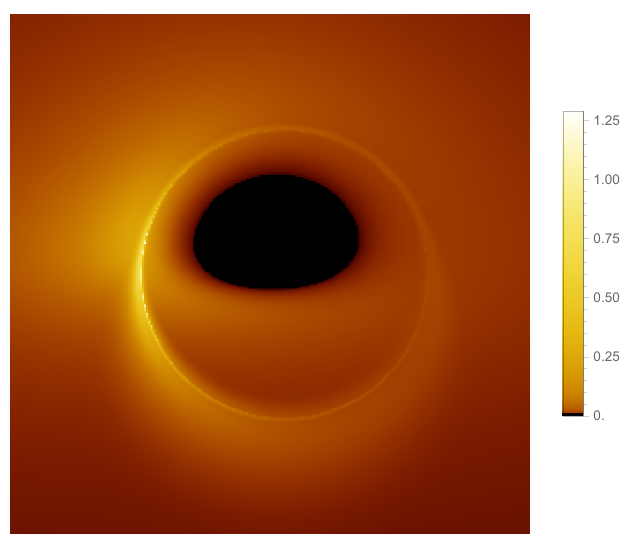}}
	
	\caption{Shadow images of rotating EGB black holes with a thin accretion disk with $M = 1$ and $a = 0.2$.}
	\label{fig1}
\end{figure}

The effect of the GB coupling constant $\xi$ on the shadow images is illustrated in Fig.~\ref{fig1}, where $\xi = 0.005$, $0.01$, and $0.015$ from top to bottom, with the spin parameter fixed at $a = 0.2$. In each image, a distinct dark region can be observed at the center, known as the inner shadow~\cite{Zhang:2023bzv}. When the strong gravitational field of the black hole causes light rays to fall directly into the event horizon, i.e., $n = 0$, Eq.~(\ref{eq:io}) gives $\tilde{I}_{o} = 0$, indicating that the corresponding pixels of the inner shadow appear completely black. Outside the inner shadow lies a luminous ring, referred to as the photon ring, which originates from light rays that cross the equatorial plane multiple times. Each intersection between a photon and the accretion disk contributes to an increase in its energy, thereby enhancing the corresponding pixel intensity. Consequently, the photon ring appears significantly brighter than the surrounding region. When the observation inclination angle is $\theta_o = 0^\circ$ (the first column), the inner shadow is perfectly circular, and the photon ring forms a symmetric annulus. This occurs because, at $\theta_o = 0^\circ$, the observer’s line of sight is perpendicular to the equatorial plane and aligned with the rotation axis of both the accretion disk and the black hole. When $\theta_o = 17^\circ$ (the second column), the inner shadow becomes slightly deformed, and the photon ring shifts toward the lower-right direction. As $\theta_o$ increases to $70^\circ$ (the third column), the inner shadow is significantly distorted into a “D”-shaped structure, and a crescent-shaped bright region emerges on the left side of the photon ring, where the brightness is markedly higher than in other regions. This phenomenon arises because a higher inclination angle amplifies the Doppler effect: in a prograde accretion disk, the light rays on the left side move toward the observer, producing a blueshift that enhances the photon energy and brightens the image, whereas the light rays on the right side move away, producing a redshift that reduces photon energy and darkens the image. Furthermore, by comparing each row, it is evident that as $\xi$ increases, the shape of the inner shadow remains unchanged, while its size gradually decreases. It is worth noting that regardless of the variations in system parameters, both the inner shadow and the photon ring persist, indicating that they are intrinsic features of the spacetime described by the metric~(\ref{eq:ma}).

\begin{figure}[!h]
	\centering 
	\subfigure[$\xi=0.005,\theta_o=0^\circ$]{\includegraphics[scale=0.4]{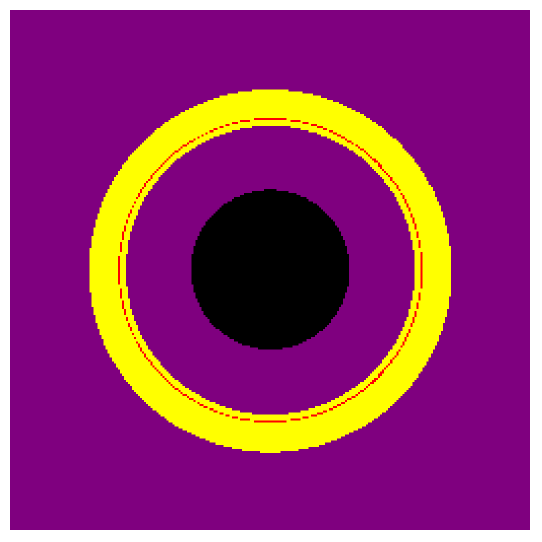}}
	\subfigure[$\xi=0.005,\theta_o=17^\circ$]{\includegraphics[scale=0.4]{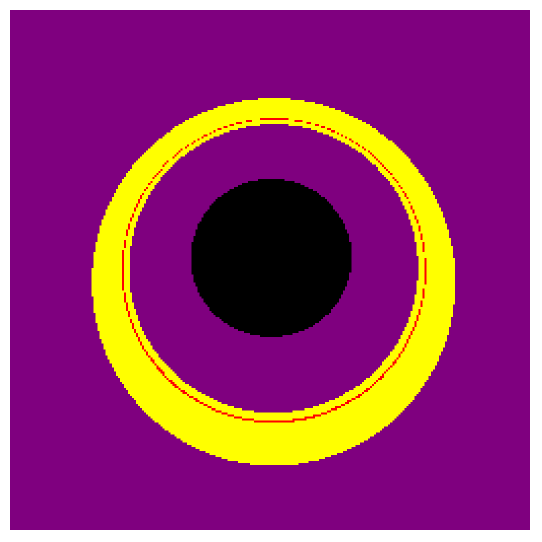}}
	\subfigure[$\xi=0.005,\theta_o=70^\circ$]{\includegraphics[scale=0.4]{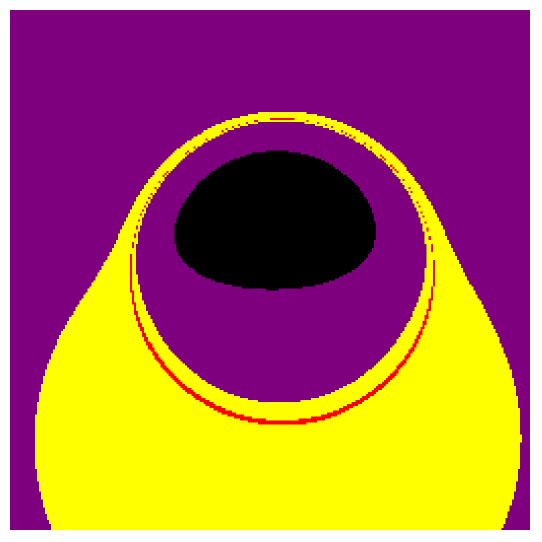}}
	
	\subfigure[$\xi=0.01,\theta_o=0^\circ$]{\includegraphics[scale=0.4]{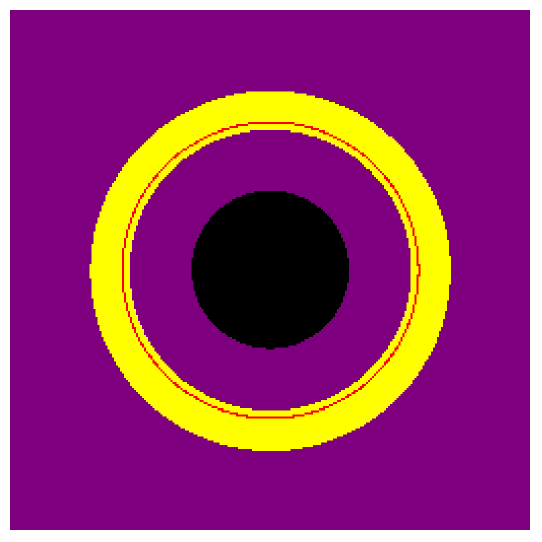}}
	\subfigure[$\xi=0.01,\theta_o=17^\circ$]{\includegraphics[scale=0.4]{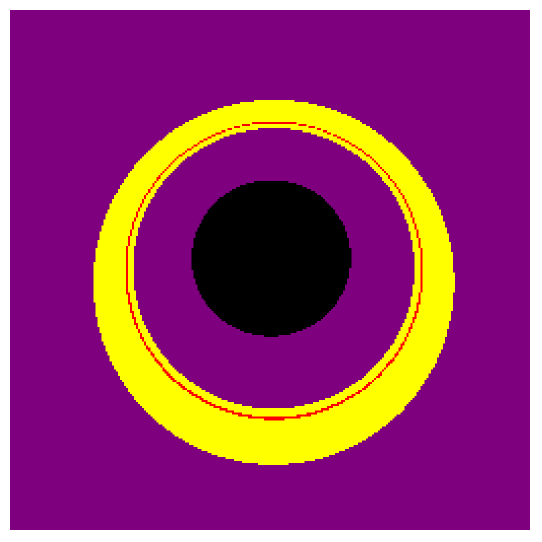}}
	\subfigure[$\xi=0.01,\theta_o=70^\circ$]{\includegraphics[scale=0.4]{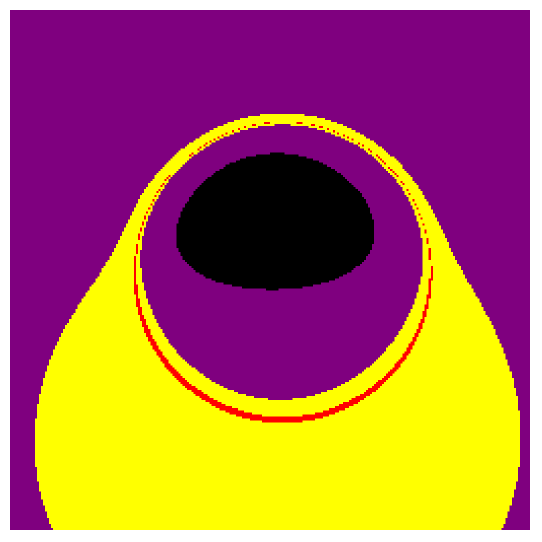}}
	
	\subfigure[$\xi=0.015,\theta_o=0^\circ$]{\includegraphics[scale=0.4]{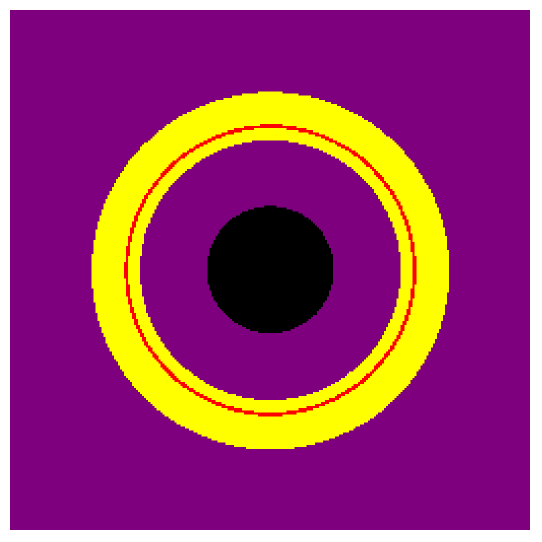}}
	\subfigure[$\xi=0.015,\theta_o=17^\circ$]{\includegraphics[scale=0.4]{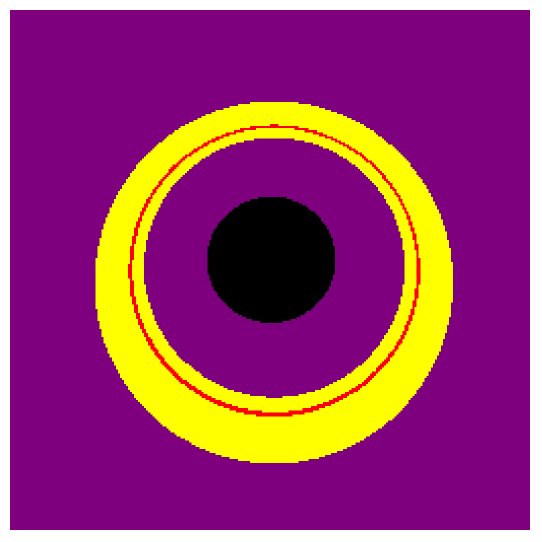}}
	\subfigure[$\xi=0.015,\theta_o=70^\circ$]{\includegraphics[scale=0.4]{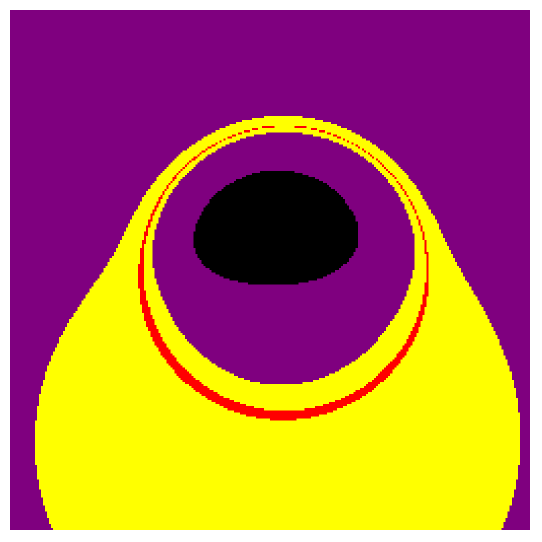}}
	
	\caption{The lensing bands of rotating EGB black holes with $a = 0.2$. The black, purple, yellow, and red regions correspond to the inner shadow, the direct image, the lensed image, and the higher-order images, respectively.}
	\label{fig2}
\end{figure}

To provide a more intuitive illustration of the number of times light rays cross the equatorial plane, Fig.~\ref{fig2} shows the lensing bands of the rotating EGB black hole, where the parameters are chosen to be consistent with those in Fig.~\ref{fig1}. In Fig.~\ref{fig2}, the black, purple, yellow, and red regions represent light rays that cross the equatorial plane zero, one, two, and more than two times, corresponding respectively to the inner shadow, the direct image, the lensed image, and the higher-order images. It can be seen that the direct image occupies the widest region, whereas the higher-order images are always contained within the area of the lensed image. When the observation inclination angle is $\theta_o = 0^\circ$ (the first column), both the lensed and higher-order images exhibit concentric ring structures that share the same center as the inner shadow. As $\xi$ increases, the ring radius slightly decreases. When $\theta_o$ increases to $17^\circ$ (the second column), the lensed and higher-order images shift downward on the screen, and the ring structures become asymmetric. When $\theta_o$ further increases to $70^\circ$ (the third column), the inner shadow, lensed image, and higher-order images are all significantly distorted. In particular, the lensed image extends markedly toward the lower part of the screen, while the higher-order images become more distinguishable in the lower region than in the upper one. The above analysis indicates that the GB coupling constant $\xi$ primarily affects the overall size of the black hole shadow, whereas the observation inclination angle $\theta_o$ governs the spatial distribution of the inner shadow and the lensed images.

\begin{figure}[!h]
	\centering 
	\subfigure[$a=0.1,\theta_o=0^\circ$]{\includegraphics[scale=0.4]{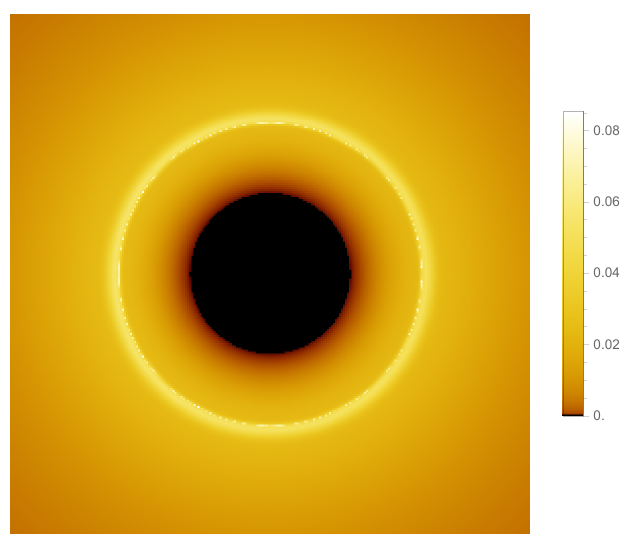}}
	\subfigure[$a=0.1,\theta_o=17^\circ$]{\includegraphics[scale=0.4]{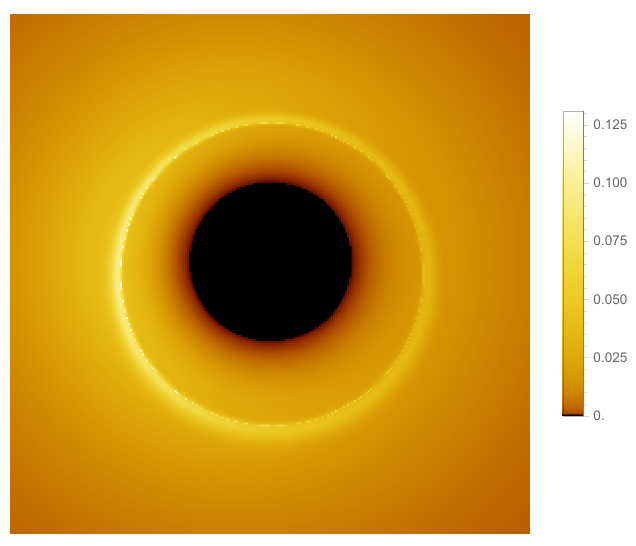}}
	\subfigure[$a=0.1,\theta_o=70^\circ$]{\includegraphics[scale=0.4]{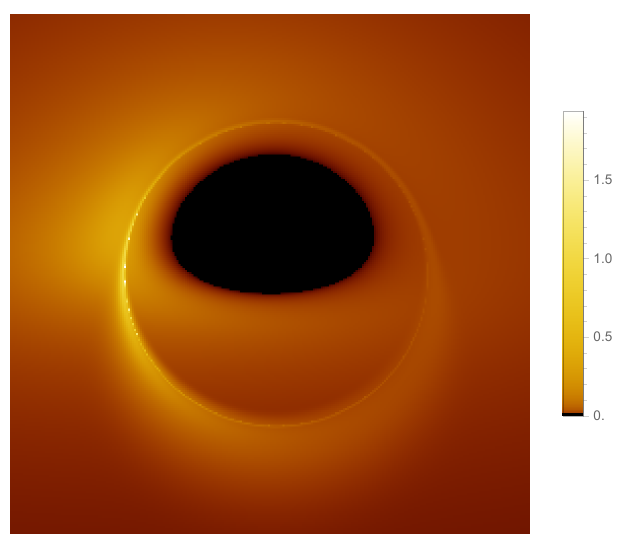}}
	
	\subfigure[$a=0.4,\theta_o=0^\circ$]{\includegraphics[scale=0.4]{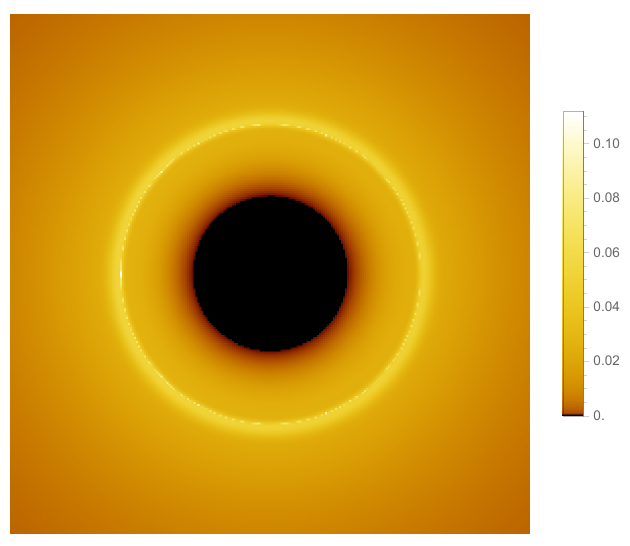}}
	\subfigure[$a=0.4,\theta_o=17^\circ$]{\includegraphics[scale=0.4]{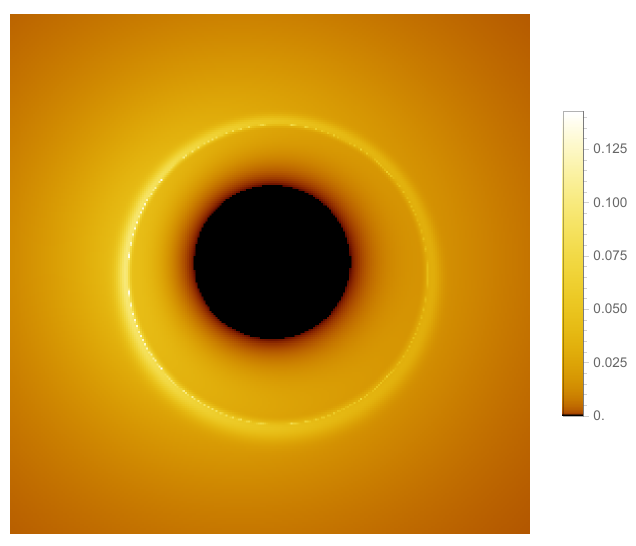}}
	\subfigure[$a=0.4,\theta_o=70^\circ$]{\includegraphics[scale=0.4]{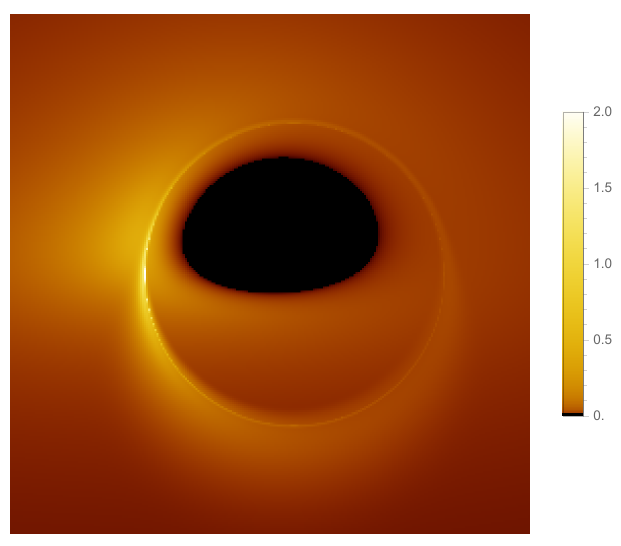}}
	
	\subfigure[$a=0.7,\theta_o=0^\circ$]{\includegraphics[scale=0.4]{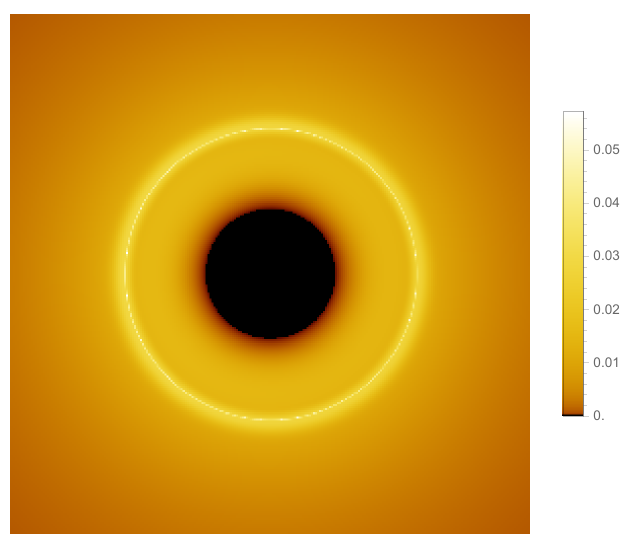}}
	\subfigure[$a=0.7,\theta_o=17^\circ$]{\includegraphics[scale=0.4]{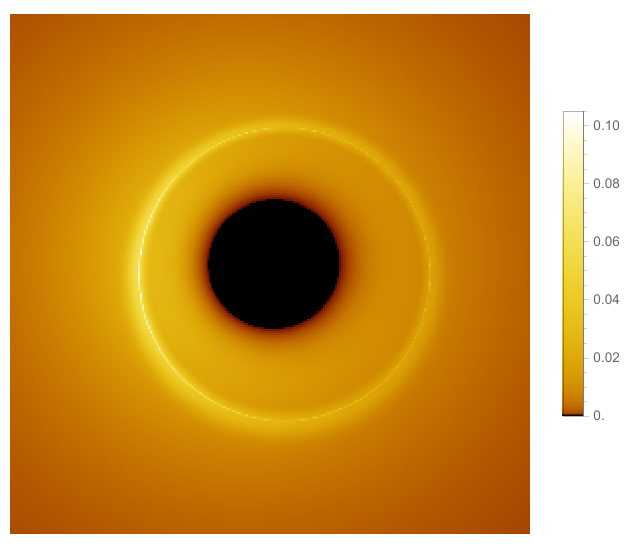}}
	\subfigure[$a=0.7,\theta_o=70^\circ$]{\includegraphics[scale=0.4]{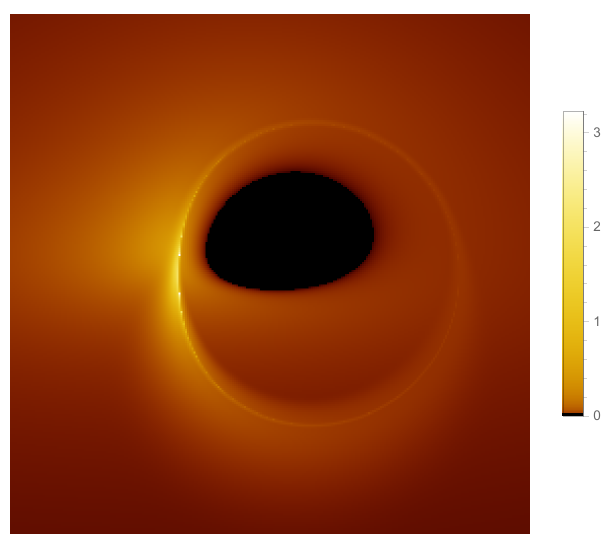}}
	
	\caption{Shadow images of rotating EGB black holes with a thin accretion disk with $M = 1$ and $\xi = 0.005$.}
	\label{fig3}
\end{figure}

\begin{figure}[!h]
	\centering 
	\subfigure[$a=0.1,\theta_o=0^\circ$]{\includegraphics[scale=0.4]{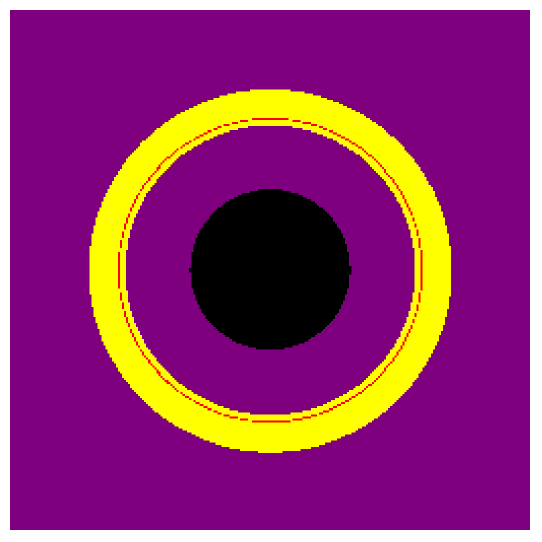}}
	\subfigure[$a=0.1,\theta_o=17^\circ$]{\includegraphics[scale=0.4]{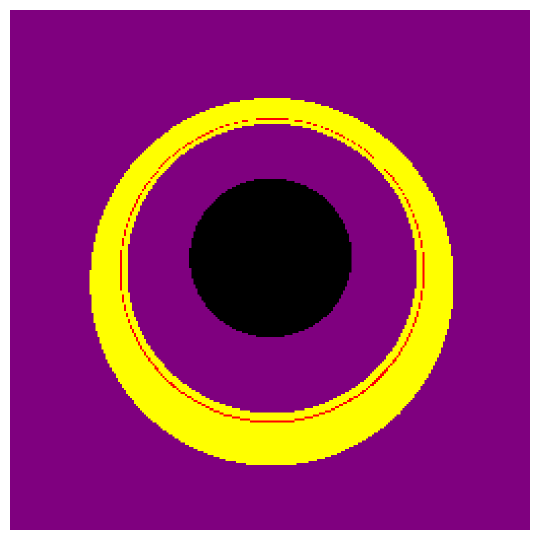}}
	\subfigure[$a=0.1,\theta_o=70^\circ$]{\includegraphics[scale=0.4]{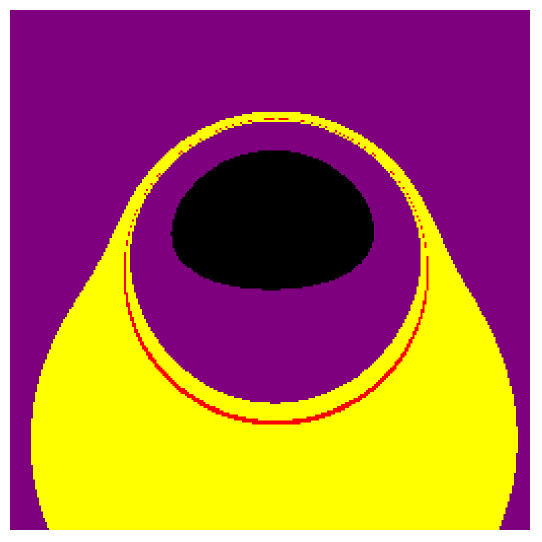}}
	
	\subfigure[$a=0.4,\theta_o=0^\circ$]{\includegraphics[scale=0.4]{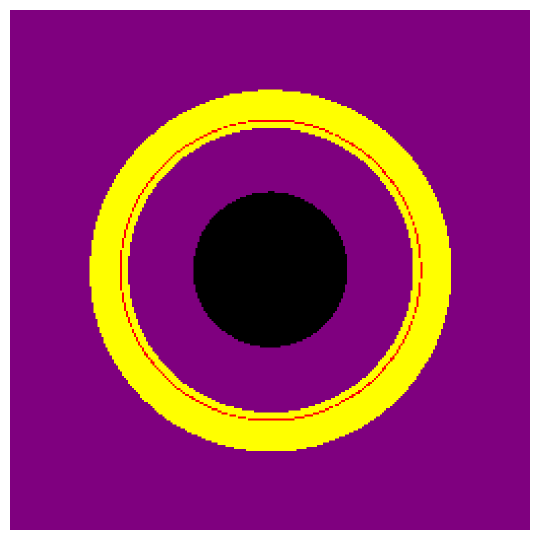}}
	\subfigure[$a=0.4,\theta_o=17^\circ$]{\includegraphics[scale=0.4]{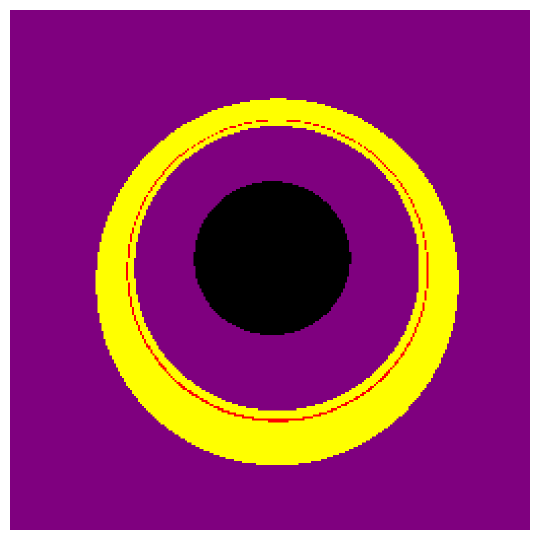}}
	\subfigure[$a=0.4,\theta_o=70^\circ$]{\includegraphics[scale=0.4]{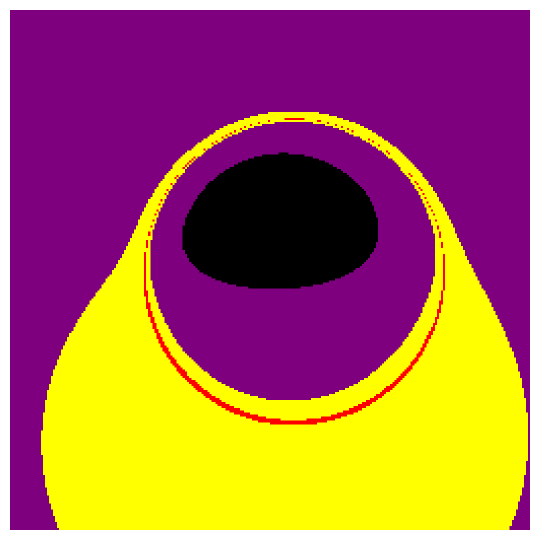}}
	
	\subfigure[$a=0.7,\theta_o=0^\circ$]{\includegraphics[scale=0.4]{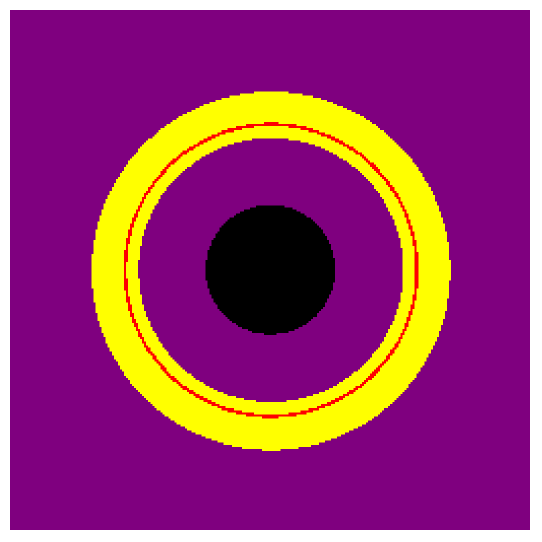}}
	\subfigure[$a=0.7,\theta_o=17^\circ$]{\includegraphics[scale=0.4]{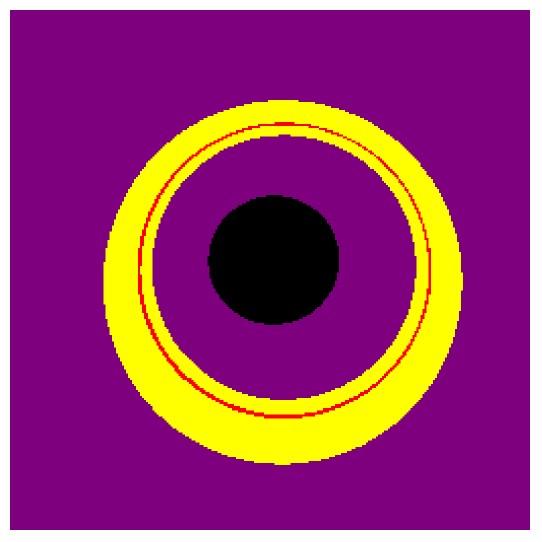}}
	\subfigure[$a=0.7,\theta_o=70^\circ$]{\includegraphics[scale=0.4]{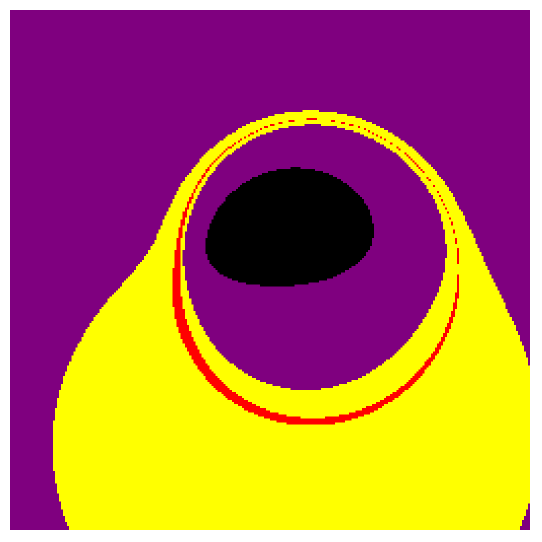}\label{fig4i}}
	
	\caption{The lensing bands of rotating EGB black holes with $\xi = 0.005$. The black, purple, yellow, and red regions correspond to the inner shadow, the direct image, the lensed image, and the higher-order images, respectively.}
	\label{fig4}
\end{figure}

Fig.~\ref{fig3} illustrates the influence of the spin parameter $a$ on the shadow images, with $\xi=0.005$ fixed. Fig.~\ref{fig4} shows the corresponding lensing bands. From Fig.~\ref{fig3}, it is evident that increasing $a$ reduces the sizes of both the inner shadow and the photon ring, while increasing $\theta_o$ leads to a deformation of the inner shadow. A noteworthy phenomenon can be observed by comparing each column of images: when $a$ and $\xi$ are fixed, changing $\theta_o$ alters the brightness symmetry of the photon ring but leaves its position nearly unchanged. For the spacetime described by the metric~(\ref{eq:ma}), once the parameters are specified, the position of the photon ring is uniquely determined by equations of motion for photons. As shown in Fig.~\ref{fig4}, the effect of $a$ on the lensing bands is similar to that of $\xi$. In the lower-left region of Fig.~\ref{fig4i}, the higher-order image zone is clearly visible, which is due to the fact that a larger spin parameter and a higher observer inclination enhance the frame-dragging effect and Doppler redshift.

\begin{figure}[!h]
	\centering 
	\subfigure[With $a = 0.2$ fixed, the red, green, and blue curves correspond to $\xi = 0.005, 0.01, 0.015$, respectively.]{\includegraphics[scale=0.8]{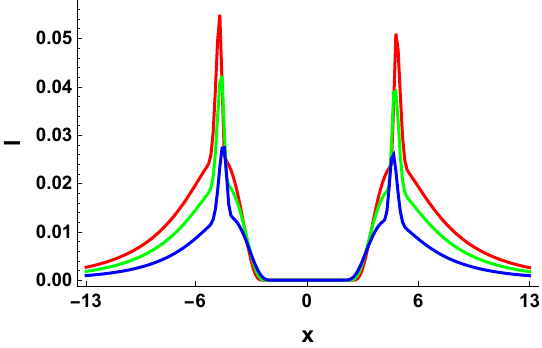}}
	\subfigure[With $\xi = 0.005$ fixed, the red, green, and blue curves correspond to $a = 0.1, 0.4, 0.7$, respectively.]{\includegraphics[scale=0.8]{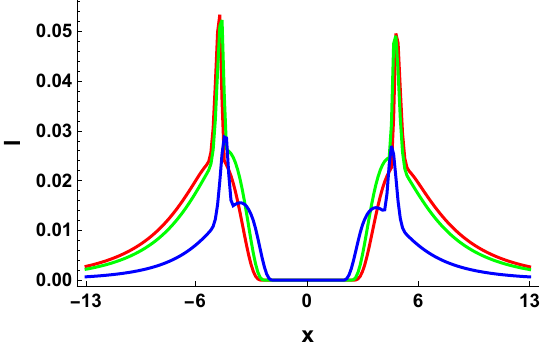}}
	
	\caption{Intensity cuts along the X-axis for $\theta_o = 0^\circ$.}	
	\label{fig5}
\end{figure}

To further analyze the effect of parameter variations on the intensity, Fig.~\ref{fig5} presents the intensity cuts along the X-axis, with $\theta_o = 0^\circ$ fixed. It can be observed that the intensity cuts are symmetrically distributed with respect to the Y-axis. Near the origin, the intensity cuts vanish, corresponding to the event horizon, while the two peaks correspond to the photon ring. As $a$ or $\xi$ increase, the peak values of the intensity cuts gradually decrease, indicating a reduction in the brightness of the photon ring. For a Schwarzschild black hole with mass set to $1$, the photon ring is located at $r_p = 3$; however, in the black hole models discussed in this work, $r_p$ is always greater than $3$, suggesting that introducing the GB coupling constant $\xi$ enlarges the radius of the photon ring.

\section{Polarization Images of EGB black hole}\label{sec5}
\subsection{The Propagation of Polarization Vector}
To gain a more comprehensive understanding of the feature of the rotating EGB spacetime, this section investigates their polarization images within the thin accretion disk model. We assume that the source of polarized radiation is synchrotron emission produced by electrons in the plasma. For an observer comoving with the plasma, whose four-velocity is $u^\mu$, the polarization direction $\vec{f}$ of the emitted light is always orthogonal to both the local magnetic field $\vec{b}$ and the photon wave vector $\vec{k}$
\begin{equation}
	\vec{f}=\frac{\vec{k}\times\vec{b}}{|\vec{k}||\vec{b}|}.
\end{equation}
The generally covariant form of the above expression can be written as
\begin{equation}
	f^\mu \propto \epsilon^{\mu\nu\alpha\beta}u_\nu k_\alpha b_\beta.
\end{equation}
During the imaging process, it is necessary to first determine the direction of the polarization vector and impose the orthonormality condition
\begin{equation}
	f^{\mu}f_{\mu}=1.
\end{equation}
The intensities of linearly polarized light and unpolarized light at the emission point are denoted by the emissivity functions $\tilde{E}_p$ and $\tilde{E}_i$, respectively. For simplicity, we assume that the emission intensity is independent of the photon frequency and magnetic field and depends only on the position. Thus,
\begin{equation}
	\tilde{E}_i=\tilde{E}_i(r),\qquad \tilde{E}_p=q \tilde{E}_i(r),
\end{equation}
where $q\in[0,1]$ describes the fraction of linearly polarized light in the total emission. If the emitted light is assumed to be fully linearly polarized, then $q=1$. Under the geometric optics approximation, the polarization vector $f^\mu$ undergoes parallel transport along the photon geodesic
\begin{equation}
	k^{\nu}\nabla_{\nu}f^{\mu}=0.
\end{equation}
This can also be expressed as
\begin{equation}
	\frac{d}{d\tau}f^{\mu}+\Gamma^{\mu}{}_{\nu\alpha}k^{\nu}f^{\alpha}=0,
\end{equation}
where $\tau$ is the affine parameter. At the observer’s position, the specific intensity of linearly polarized light $\tilde{P}_{\nu_o}$ and the total intensity $\tilde{I}_{\nu_o}$ are given by the same form as in the unpolarized case
\begin{equation}
	\tilde{P}_{\nu_o}=\chi^3\tilde{E}_p,\qquad \tilde{I}_{\nu_o}=\chi^3\tilde{E}_i,
\end{equation}
where $\chi$ is the redshift factor. In Eq.~(\ref{e32}), we construct a set of orthogonal tetrads ZAMO at the observer’s position and define the imaging screen accordingly. Choosing an orthonormal basis $(e_{(\theta)},\,e_{(\varphi)})$ on the screen, the projection of the polarization vector onto the image plane is given by
\begin{equation}
	f^{(\alpha)}=f^\mu\cdot e_\alpha=-f^\mu\cdot e_\varphi,\qquad f^{(\beta)}=f^\mu\cdot e_\beta=-f^\mu\cdot e_\theta.
\end{equation}
After determining the polarization vector direction, the total linearly polarized intensity can be obtained by summing the contributions from each emission point in the equatorial plane. The Stokes parameters $Q$ and $U$ satisfy the principle of linear superposition~\cite{Huang:2024bar}. Therefore, the final results are obtained by summing $Q$ and $U$
\begin{equation}
	Q_{all}=\sum_{n=1}^{N}\chi_n^3\tilde{E}_{pn}\left[\left(f_n^{(\alpha)}\right)^2-\left(f_n^{(\beta)}\right)^2\right],\quad
	U_{all}=\sum_{n=1}^{N}\chi_n^3\tilde{E}_{pn}\left(2f_n^{(\alpha)}f_n^{(\beta)}\right).
\end{equation}
Based on the above analysis, the total linearly polarized intensity and the electric vector position angle (EVPA) of the observed light can be calculated as
\begin{equation}
	\tilde{P}_o=\left(Q_{all}^2+U_{all}^2\right)^{\frac12},\quad\Theta_E=\frac{1}{2}\arctan\frac{U_{all}}{Q_{all}}.
\end{equation}
After obtaining the EVPA and combining it with the polarization intensity, the polarization image of a rotating EGB black hole in the thin accretion disk background can be constructed.

\subsection{Numerical Results}

\begin{figure}[!h]
	\centering 
	\subfigure[$\xi=0.005,\theta_o=0^\circ$]{\includegraphics[scale=0.4]{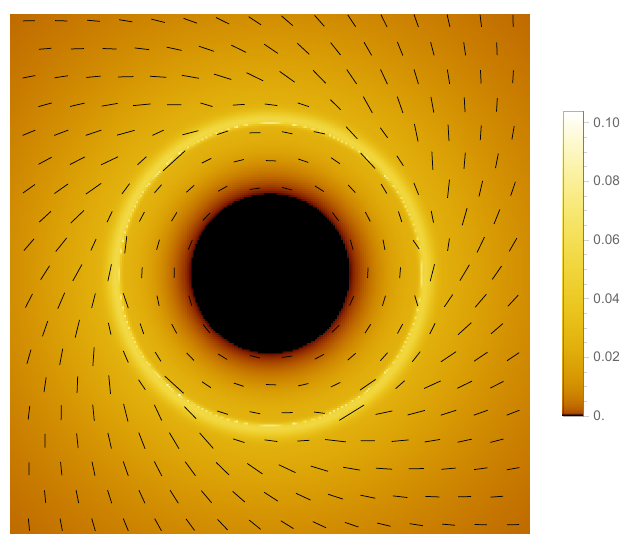}}
	\subfigure[$\xi=0.005,\theta_o=17^\circ$]{\includegraphics[scale=0.4]{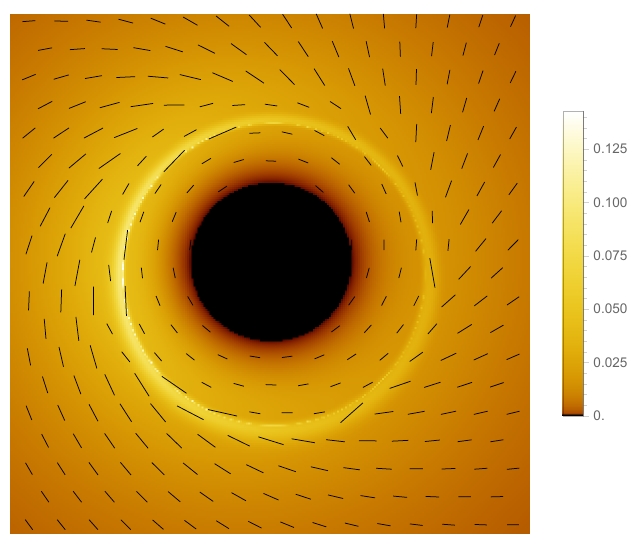}}
	\subfigure[$\xi=0.005,\theta_o=70^\circ$]{\includegraphics[scale=0.4]{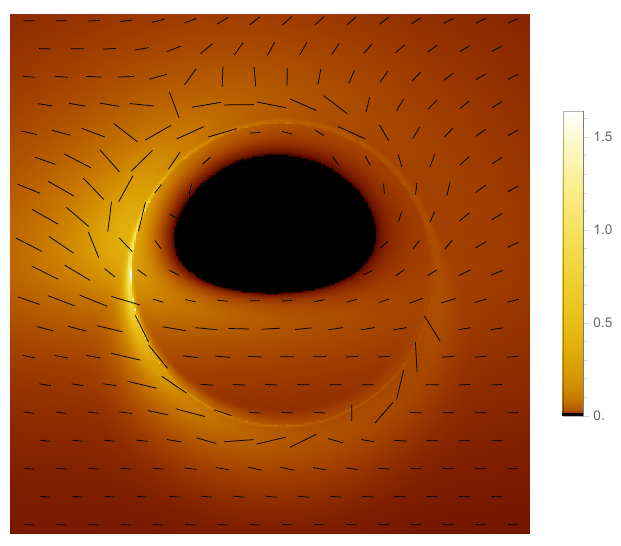}}
	
	\subfigure[$\xi=0.01,\theta_o=0^\circ$]{\includegraphics[scale=0.4]{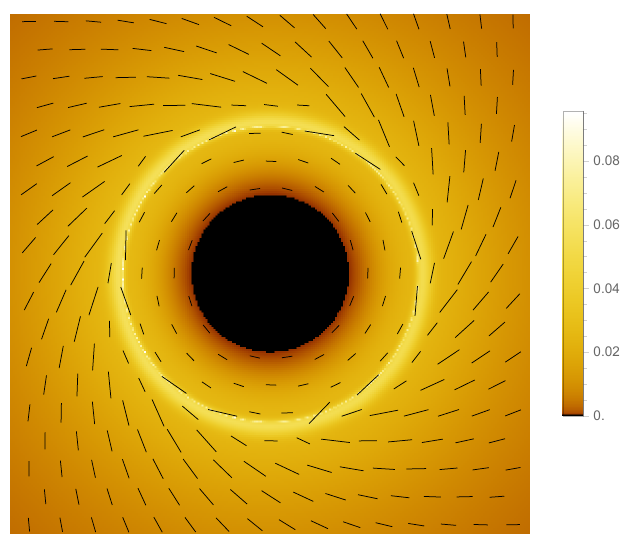}}
	\subfigure[$\xi=0.01,\theta_o=17^\circ$]{\includegraphics[scale=0.4]{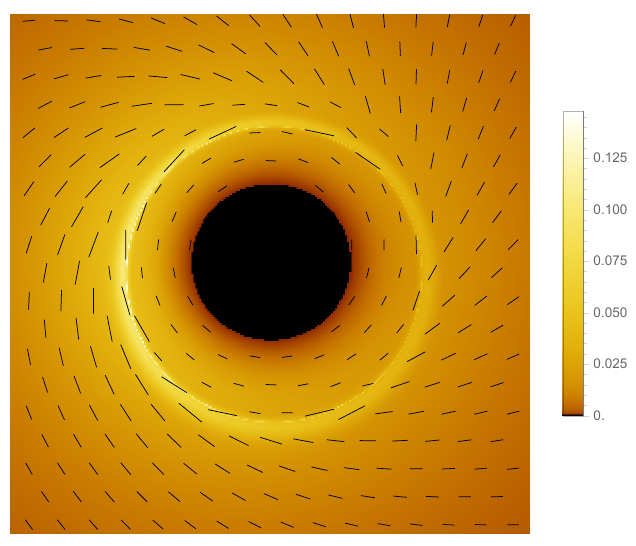}}
	\subfigure[$\xi=0.01,\theta_o=70^\circ$]{\includegraphics[scale=0.4]{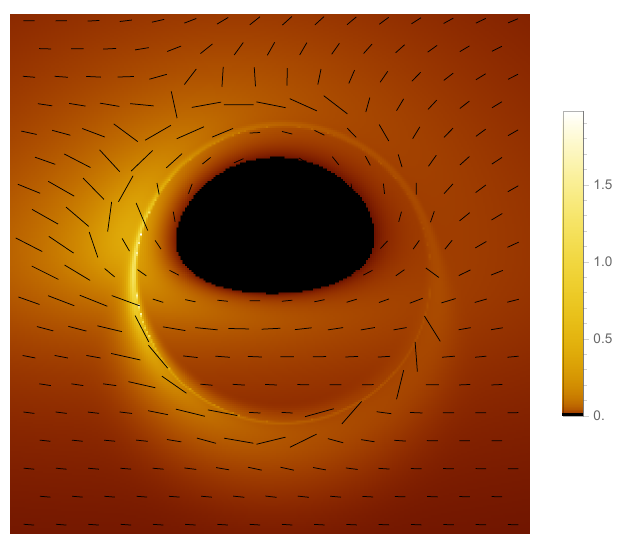}}
	
	\subfigure[$\xi=0.015,\theta_o=0^\circ$]{\includegraphics[scale=0.4]{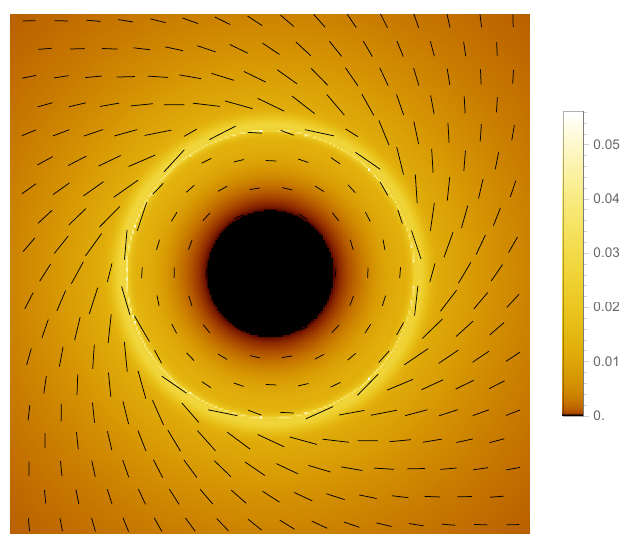}}
	\subfigure[$\xi=0.015,\theta_o=17^\circ$]{\includegraphics[scale=0.4]{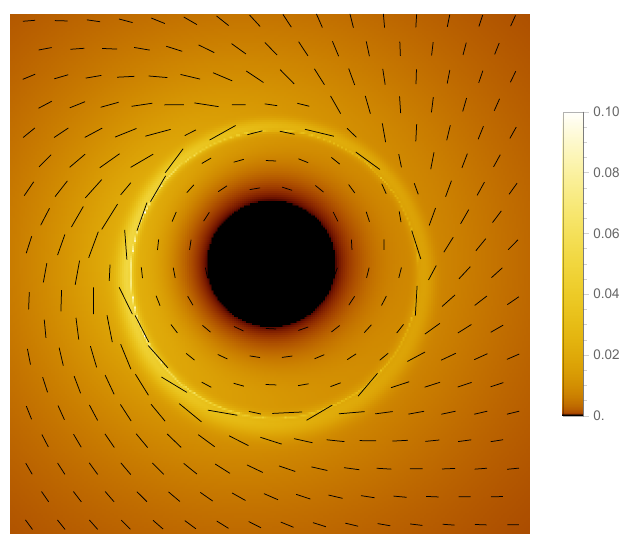}}
	\subfigure[$\xi=0.015,\theta_o=70^\circ$]{\includegraphics[scale=0.4]{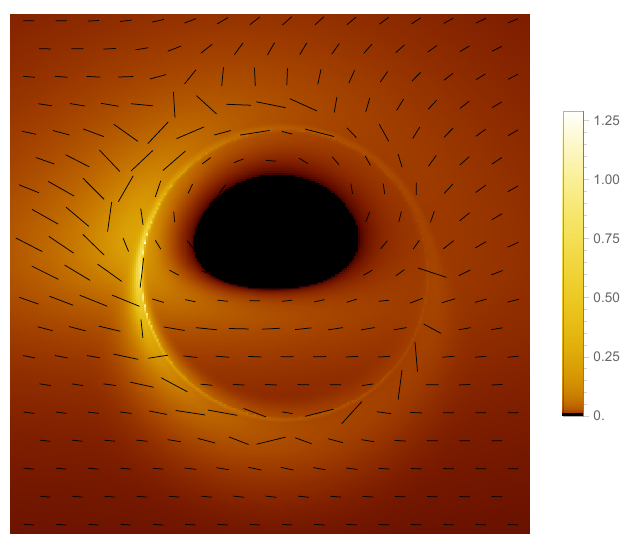}}
	
	\caption{Polarization images of rotating EGB black holes with $a = 0.2$.}
	\label{fig6}
\end{figure}

\begin{figure}[!h]
	\centering 
	\subfigure[$a=0.1,\theta_o=0^\circ$]{\includegraphics[scale=0.4]{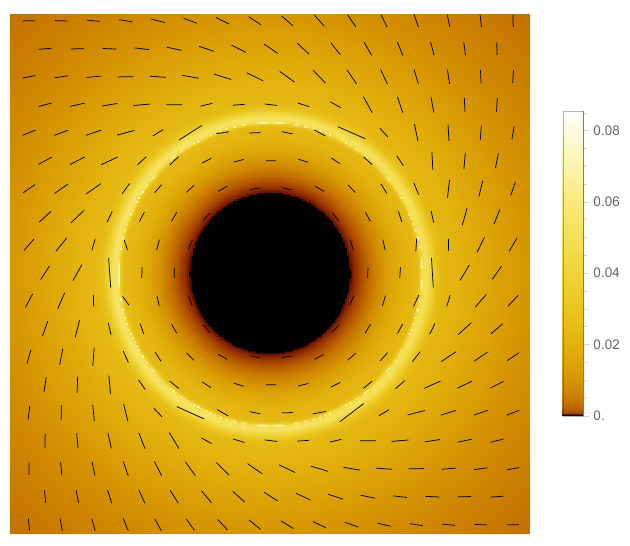}}
	\subfigure[$a=0.1,\theta_o=17^\circ$]{\includegraphics[scale=0.4]{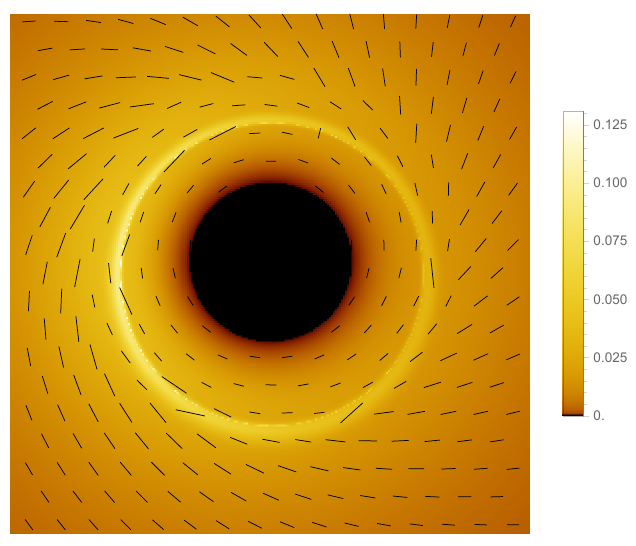}}
	\subfigure[$a=0.1,\theta_o=70^\circ$]{\includegraphics[scale=0.4]{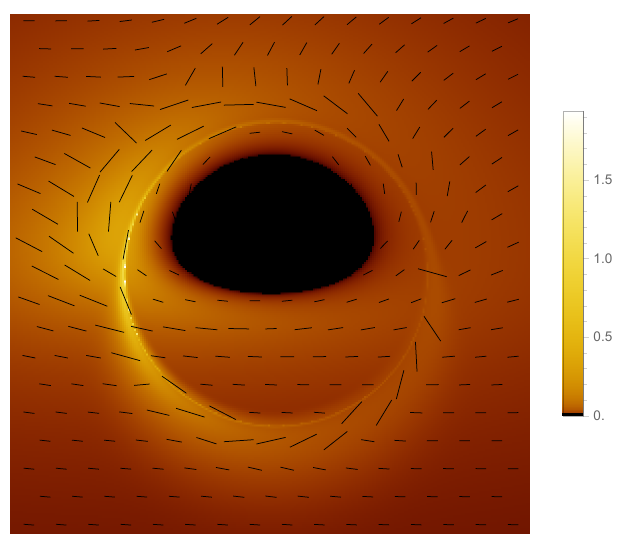}}
	
	\subfigure[$a=0.4,\theta_o=0^\circ$]{\includegraphics[scale=0.4]{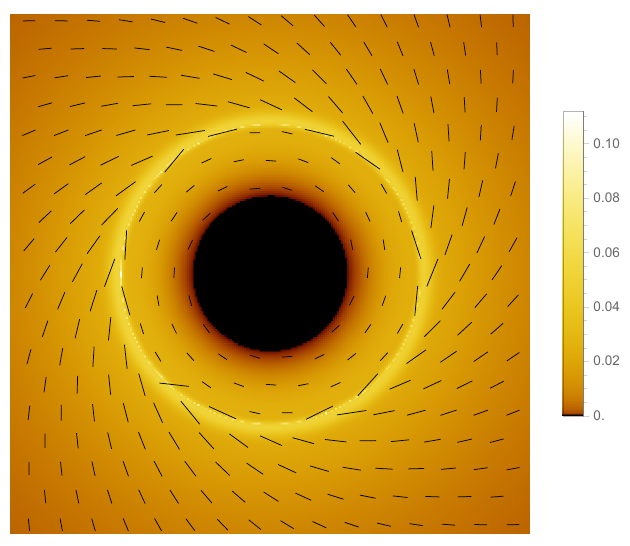}}
	\subfigure[$a=0.4,\theta_o=17^\circ$]{\includegraphics[scale=0.4]{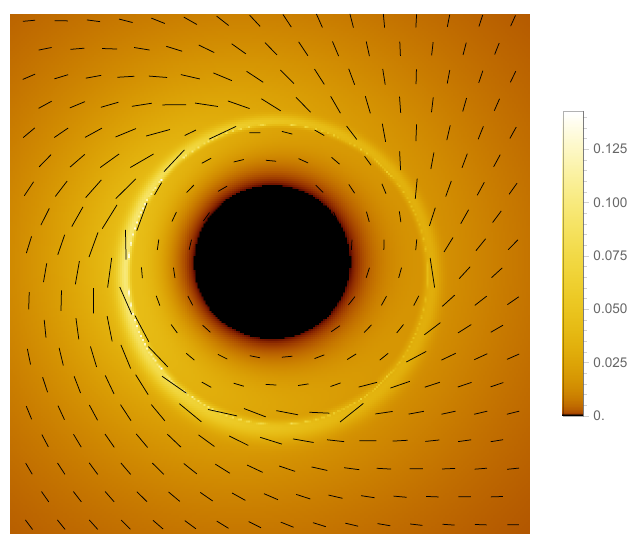}}
	\subfigure[$a=0.4,\theta_o=70^\circ$]{\includegraphics[scale=0.4]{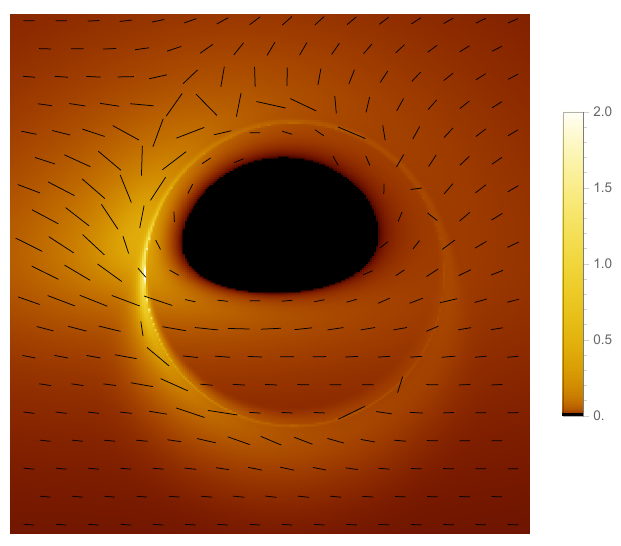}}
	
	\subfigure[$a=0.7,\theta_o=0^\circ$]{\includegraphics[scale=0.4]{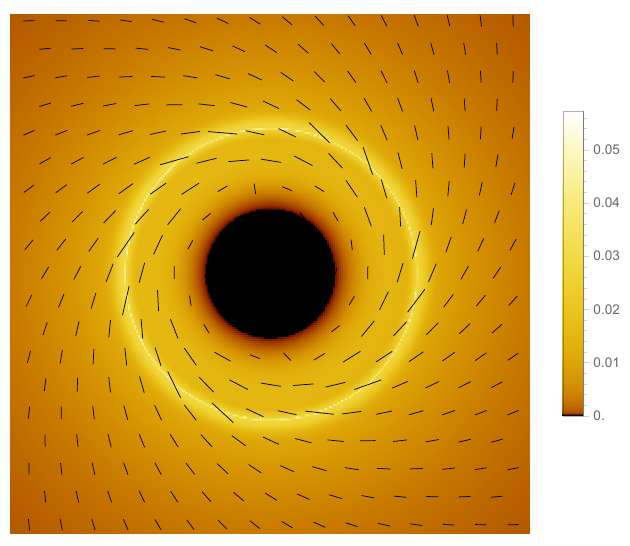}}
	\subfigure[$a=0.7,\theta_o=17^\circ$]{\includegraphics[scale=0.4]{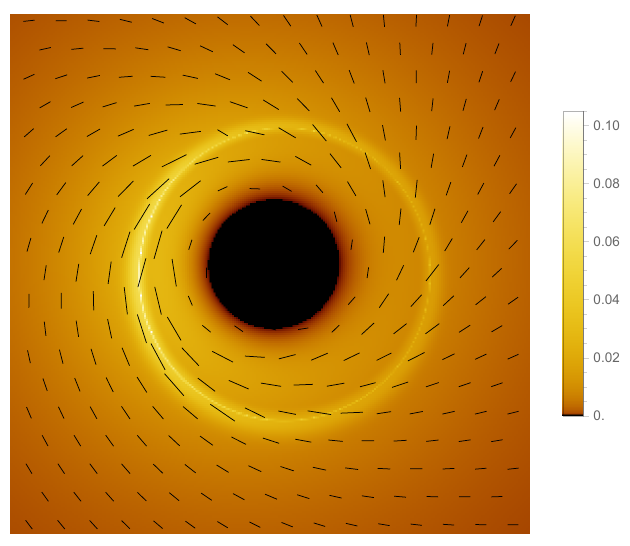}}
	\subfigure[$a=0.7,\theta_o=70^\circ$]{\includegraphics[scale=0.4]{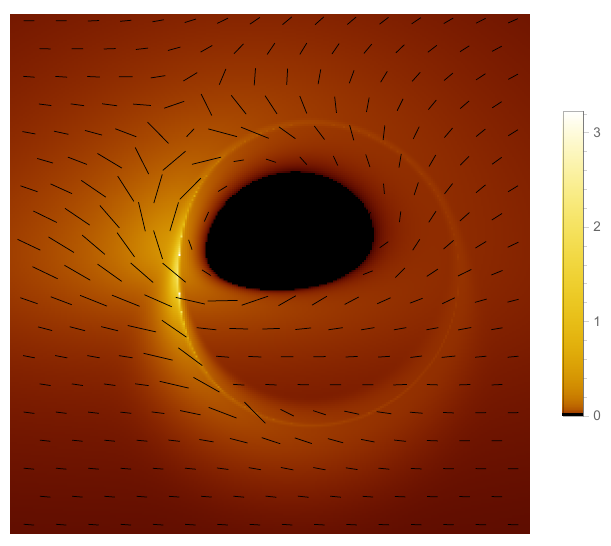}}
	
	\caption{Polarization images of rotating EGB black holes with $\xi = 0.005$.}
	\label{fig7}
\end{figure}

For M87*, assuming it to be a Schwarzschild black hole, it has been demonstrated that the optimal magnetic field configuration $\vec{b} = (0.87, 0.5, 0)$ can successfully reproduce its observed polarization features. Therefore, the same $\vec{b}$ is adopted in our simulations. Similar to the previous analysis, we consider three observer inclination angles: $\theta_o = 0^\circ, 17^\circ, 70^\circ$. In this consideration, the polarization images are presented in Fig.~\ref{fig6} and~\ref{fig7}.

Fig.~\ref{fig6} and~\ref{fig7} show the effects of $\xi$ and $a$, respectively, on the polarization vectors. In each figure, the length and direction of the black arrows represent the linearly polarized intensity $\tilde{P}_o$ and the electric vector position angle $\Theta_E$, while the background image corresponds to the thin accretion disk emission, consistent with Figs.~\ref{fig1} and~\ref{fig3}. It can be seen that the magnitude of $\tilde{P}_o$ is positively correlated with the total intensity: the linearly polarized intensity is significantly stronger in bright regions than in dim regions and reaches its maximum near the photon ring. Notably, when $\theta_o$ is small, the rotation of the polarization vectors appears more ordered compared to the case with a larger observation inclination angle. 
Meanwhile, as the parameter $\xi$ increases, the polarization direction near the inner shadow changes significantly, indicating that the polarization direction can also serve as a probe for testing the validity of the EGB gravity theory.
It should also be noted that, for black holes, radiation cannot escape beyond the event horizon; theoretically, no polarization effects can be observed within this region. However, for horizonless compact objects (such as boson stars), strong polarization signatures can appear in the central region~\cite{Li:2025awg,Zeng:2025fjg}.

\section{Conclusion and Discussion}\label{sec6}

In this work, we systematically investigated the shadow and polarization images of rotating EGB black holes based on the ray-tracing method. For the shadow images, we considered a geometrically and optically thin accretion disk located in the equatorial plane as the light source; for the polarization images, we assumed that the polarized emission originates from synchrotron radiation produced by electrons in the plasma. On this basis, we focused on analyzing the effects of the GB coupling constant $\xi$, the spin parameter $a$, and the observation inclination angle $\theta_o$ on both types of images.

For the thin disk images, we find that regardless of how the parameter space $(\xi, a, \theta_o)$ varies, the image always exhibits a central dark region and a bright ring, corresponding to the inner shadow and the photon ring, respectively. As $\theta_o$ increases, the inner shadow deforms significantly from a circular shape to a characteristic ``D'' shape. A larger $\theta_o$ also enhances the Doppler redshift, causing the brightness of the photon ring to become asymmetric. In the case of a prograde accretion disk, a crescent-shaped bright region appears on the left side of the photon ring. However, changes in $\theta_o$ do not affect the position of the photon ring. These results indicate that the inner shadow and the photon ring are intrinsic features of the spacetime of rotating EGB black holes. On the other hand, the size of the inner shadow is highly sensitive to variations in $\xi$ and $a$. Increasing either $\xi$ or $a$ reduces the size of the inner shadow but does not alter its overall shape. In addition, we further studied the lensing bands of rotating EGB black holes, where different colors represent the number of times light rays cross the equatorial plane. The analysis shows that higher-order images are always located within the region of the lensing image. When $\theta_o = 0^\circ$, the inner shadow, lensing image, and higher-order images form concentric rings. Increasing $\theta_o$ causes the lensing image to shift downward on the screen, and the higher-order images in the lower part of the screen become significantly more distinguishable than those in the upper part. If the spin parameter $a$ is further increased under these conditions, the frame-dragging effect enhances the visibility of the higher-order images in the lower part of the screen. Meanwhile, it shows that the GB parameter increase the region of lensing bands.
Finally, based on the intensity cuts along the X-axis for $\theta_o = 0^\circ$, the GB coupling constant $\xi$ does not only decrease the radius of the photon ring, but the radius of inner shadow, and its reducing effect on the inner shadow is more pronounced. And, the maximum value of the intensity also decreases accordingly with $\xi$.

For the polarization images, we find that the polarization intensity is positively correlated with the brightness of the optical image; the polarization strength in high-brightness regions is significantly greater than in low-brightness regions. Each time a photon crosses the equatorial plane, its intensity increases, causing the polarization intensity to reach its maximum near the photon ring. Both the magnitude and orientation of the polarization exhibit strong dependence on $(\xi, a, \theta_o)$, indicating that the polarization features of rotating EGB black holes can effectively reflect the intrinsic spacetime structure. Interestingly, due to the presence of the event horizon, no polarization effects can be observed within the inner shadow region. This provides a potential criterion for distinguishing black holes from horizonless compact objects, such as boson stars and neutron stars.

Our study reveals the impact of the GB parameter $\xi$ on black hole images within the thin accretion disk model. Compared with relying solely on accretion disk image, combining accretion disk images with polarization effects allows for a more comprehensive and in-depth characterization of the unique spacetime structure inherent to EGB black holes. In future work, we plan to consider more realistic accretion disk models, such as analytic thick disks, and to further perform numerical simulations of EGB black hole shadow images. These efforts are expected to provide valuable theoretical guidance for future astronomical observations and serve as a powerful tool for the observational testing of EGB gravity.


\cleardoublepage

\vspace{10pt}
\noindent {\bf Acknowledgments}

\noindent
This work is supported by the National Natural Science Foundation of China (GrantNo.12505059).

\bibliographystyle{JHEP} 
\bibliography{biblio} 

\end{document}